\documentclass[journal]{IEEEtran}
\usepackage[normalem]{ulem}
\usepackage[noadjust]{cite}
\usepackage{bbm}
\usepackage{array}
\usepackage{dcolumn}
\usepackage{epsfig}
\usepackage[intlimits]{amsmath}
\usepackage{amsmath, amsfonts, yhmath, bm}
\usepackage{amssymb}
\usepackage{psfrag}
\usepackage{color,soul}
\usepackage{xcolor,cancel}
\usepackage[normalem]{ulem}
\usepackage{enumerate}
\usepackage{stackengine}
\usepackage[noadjust]{cite}
\usepackage{graphicx}
\usepackage{caption}
\usepackage{subcaption}
\usepackage{multirow}
\usepackage{cite}
\usepackage[font=footnotesize]{caption}
\usepackage{etoolbox}
\usepackage{tcolorbox}
\usepackage{float}
\newcommand {\myvec}[1] {{\mbox{\boldmath $#1$}}}
\newcommand {\mymat}[1]  {{\mbox{\boldmath $#1$}}}
\newcommand{\bs}[1]{\boldsymbol{#1}}

\DeclareMathAlphabet      {\mathbfit}{OML}{cmm}{b}{it}

\DeclareRobustCommand{\rchi}{{\mathpalette\irchi\relax}}
\newcommand{\irchi}[2]{\raisebox{\depth}{$#1\chi$}} 

\newcommand{\etal}{\textit{et al.}}

\newcommand {\mS} {\mymat{S}}
\newcommand {\A} {\mymat{A}}

\newcommand {\hC} {\widehat{\C}}
\newcommand {\meps} {\mymat{\mathcal{E}}}

\newcommand {\mLambda} {\mymat{\Lambda}}
\newcommand {\hA} {\widehat{\A}}

\newcommand {\hS} {\widehat{\mS}}

\newcommand {\tX} {\widetilde{\X}}
\newcommand {\tV} {\widetilde{\V}}
\newcommand {\tZ} {\widetilde{\Z}}
\newcommand {\tS} {\widetilde{\mS}}
\newcommand {\C} {\mymat{C}}

\newcommand {\Z} {\mymat{Z}}

\newcommand {\E} {\mymat{E}}

\newcommand {\Ep} {\mymat{\mathcal{E}}}

\newcommand {\tC} {\widetilde{\C}}

\renewcommand {\P} {\mymat{P}}
\newcommand {\tP} {\widetilde{\Q}}

\newcommand {\Q} {\mymat{Q}}

\newcommand {\htC} {\widehat{\tC}}
\newcommand {\htP} {\widehat{\tP}}

\newcommand {\I} {\mymat{I}}
\newcommand {\X} {\mymat{X}}

\newcommand {\V} {\mymat{V}}
\newcommand {\F} {\mymat{F}}

\newcommand {\mChi} {\mymat{\rchi}}
\newcommand {\ue} {\myvec{e}}
\newcommand {\ua} {\myvec{a}}

\newcommand {\uxi} {\myvec{\xi}}
\newcommand {\huxi} {\widehat{\myvec{\xi}}}

\newcommand {\uepsilon} {\myvec{\varepsilon}}

\newcommand {\ux} {\myvec{x}}
\newcommand {\ud} {\myvec{d}}
\newcommand {\utx} {\widetilde{\myvec{x}}}
\newcommand {\uts} {\widetilde{\myvec{s}}}
\newcommand {\utv} {\widetilde{\myvec{v}}}
\newcommand {\utz} {\widetilde{\myvec{z}}}

\newcommand {\uv} {\myvec{v}}

\newcommand {\uo} {\myvec{0}}
\newcommand {\us} {\myvec{s}}
\newcommand {\ulambda} {\myvec{\lambda}}
\newcommand {\hulambda} {\widehat{\myvec{\lambda}}}

\newcommand {\uf} {\myvec{f}}

\newcommand {\utheta} {\myvec{\theta}}
\newcommand {\hutheta} {\widehat{\myvec{\theta}}}

\newcommand {\Rset} {\mathbb{R}}
\newcommand {\Cset} {\mathbb{C}}
\newcommand {\Zset} {\mathbb{Z}}
\newcommand {\Eset} {\mathbb{E}}

\newcommand {\Tr} {\text{\normalfont Tr}}
\newcommand {\tps} {\rm{T}}

\begin{document}

\title{A Maximum Likelihood-Based Minimum Mean Square Error Separation and Estimation of Stationary Gaussian Sources from Noisy Mixtures}

\author{Amir Weiss and Arie Yeredor \\
	School of Electrical Engineering, Faculty of Engineering, Tel-Aviv University \\
	\{amirwei2@mail,arie@eng\}.tau.ac.il
	
}

\maketitle

\begin{abstract}
In the context of Independent Component Analysis (ICA), noisy mixtures pose a dilemma regarding the desired objective. On one hand, a ``maximally separating" solution, providing the minimal attainable Interference-to-Source-Ratio (ISR), would often suffer from significant residual noise. On the other hand, optimal Minimum Mean Square Error (MMSE) estimation would yield estimates which are the ``closest possible" to the true sources, often at the cost of compromised ISR. In this work, we consider noisy mixtures of temporally-diverse stationary Gaussian sources in a semi-blind scenario, which conveniently lends itself to either one of these objectives. We begin by deriving the ML Estimates (MLEs) of the unknown (deterministic) parameters of the model: the mixing matrix and the (possibly different) noise variances in each sensor. We derive the likelihood equations for these parameters, as well as the corresponding Cram\'er-Rao lower bound, and propose an iterative solution for obtaining the MLEs. Based on these MLEs, the asymptotically-optimal ``maximally separating" solution can be readily obtained. However, we also present the ML-based MMSE estimate of the sources, alongside a frequency-domain-based computationally efficient scheme, exploiting their stationarity. We show that this estimate is asymptotically optimal and attains the (oracle) MMSE lower bound. Furthermore, for non-Gaussian signals, we show that this estimate serves as a Quasi ML (QML)-based Linear MMSE (LMMSE) estimate, and attains the (oracle) LMMSE lower bound asymptotically. Empirical results of three simulation experiments are presented, corroborating our analytical derivations.
\end{abstract}

\begin{IEEEkeywords}
Semi-blind source separation, independent component analysis, maximum likelihood, minimum mean square error, Cram\'er-Rao lower bound.
\end{IEEEkeywords}
\vspace{-0.3cm}
\section{Introduction}
Blind Source Separation (BSS) \cite{cardoso1990eigen,jutten1991blind,comon1991blind} is the problem of retrieving a (known) number of signals of interest, termed the sources, from a number of mixed versions thereof, termed the mixtures. In classical Independent Component Analysis (ICA) \cite{comon2010handbook,comon1992independent,hyvarinen2004independent}, one of the most popular approaches for BSS, the mixtures are assumed to be linear combinations of mutually statistically independent sources. The term ``blind" refers to the fact that no further prior knowledge is available.

However, in some cases, commonly referred to as ``semi-blind" \cite{gunther2012learning,hesse2006semi,nesta2011batch}, some \textit{a-priori} statistical (full/partial) information on the sources is available. A special case is when the sources' probability distributions are known, thus allowing the Maximum Likelihood (ML) approach \cite{pham1997blind,pearlmutter1997maximum,cardoso1997infomax,jang2003maximum,degerine2004separation,yeredor2011empirical} to be taken. It is well known that the ML approach leads (under mild conditions) to asymptotic optimality \cite{cramer2016mathematical}, in the sense of Minimum Mean Square Error (MMSE), in unbiased estimation of the unknown (deterministic) parameters of the underlying model - the mixing matrix elements and (possibly) other parameters. This, in turn, leads to (asymptotically) optimal unbiased separation \cite{koldovsky2005cramer,doron2007cramer,yeredor2010blind} in the sense of minimal Interference-to-Source Ratio (ISR), a common separation measure widely used in the context of BSS with a noise-free model. For this noiseless model, the ML Estimate (MLE) of the demixing matrix enjoys the \textit{equivariance} property (e.g., \cite{cardoso1996equivariant,yeredor2010blind}). In the context of ICA, a separation approach is said to be ``equivariant in the mixing matrix" if its resulting ISR does not depend on the true value of the mixing matrix but only on the sources' statistics. However, this appealing property, which is shared by many (but not by all) ICA algorithms, holds true only for the noise-free model.

In the more realistic case, the received signals are some noisy versions of the mixtures, where additive noise, statistically independent of the sources, is often a suitable model for describing the noisy mixtures. Interestingly, the noisy case has apparently seen less treatment than its noiseless counterpart. Cardoso and Souloumiac presented the Joint Approximate Diagonalization of Eigen-matrices (JADE) algorithm for separation of non-Gaussian sources, based on 4-th order sample cumulants of the mixtures, taking into account possible additive Gaussian noise. In \cite{belouchrani1994maximum}, Belouchrani and Cardoso took the ML approach (using the Expectation-Maximization (EM) algorithm) in a semi-blind scenario, where each source transmits symbols from a known alphabet with known \textit{a-priori} probabilities, incorporating additive temporally-white Gaussian noise. Moulines \etal{} \cite{moulines1997maximum} also presented an ML approach based on the EM algorithm for noisy mixtures, but modeled the sources' distributions as mixtures of Gaussians, and considered both instantaneous and convolutive mixtures. The Second-Order Blind Identification (SOBI) algorithm for noisy instantaneous mixtures was presented in \cite{belouchrani1997blind} by Belouchrani \etal{}, exploiting the time coherence of stationary sources, based only on Second-Order Statistics (SOS). SOBI enables the separation of mixtures of Gaussian sources (previously considered despicable in the context of ICA) under the SOS identifiability condition \cite{belouchrani1997blind}. Of course, noisy mixtures have also been considered in other various scenarios as well (e.g., \cite{lee1999blind,sahmoudi2005blind}).

Indeed, once Gaussian (temporally-diverse) sources have first been considered in the context of ICA, quite a few fundamental results have been achieved for the noise-free model (see, e.g., \cite{yeredor2010blind} and reference therein), which in the Gaussian case conveniently lends itself to tractable, asymptotically optimal ML separation. These are informative and also approximately valid in ``slightly noisy" models for the high Signal-to-Noise-Ratio (SNR) regime. However, to the best of our knowledge, an optimal (or asymptotically optimal) separation-estimation scheme of Gaussian sources from \textit{noisy} mixtures, where equivariance does not hold, has not yet been proposed. Therefore, in this paper, it is our purpose to address this problem and propose such an asymptotically optimal separation-estimation scheme. More specifically, we consider the semi-blind scenario, where the received signals are known (or presumed) to be noise-contaminated linear mixtures of temporally-diverse, stationary (real-valued) Gaussian signals with known, distinct spectra\footnote{By ``distinct spectra" we mean that no spectrum of any of the sources is a scaled version of the spectrum of another source.}.

Our proposed solution sets a theoretical benchmark of the best asymptotically attainable performance for this model, in terms of both separation and estimation of the sources, which serves as our main motivation for this work. Additionally, our proposed model is suitable for mixtures which arise, for example, in Visible Light Communication (VLC) systems involving Multiple-Input Multiple-Output (MIMO) transmission schemes (e.g., \cite{zeng2009high,nuwanpriya2015indoor,xu2017channel}), commonly used for attaining higher transmission rates and/or enhanced (post-processing) SNR. Since such systems (and others) are operated in various conditions, their performance is usually evaluated in many operation modes. In particular, overall performance measures (such as Bit-Error-Rate (BER)) are evaluated in a wide range of SNRs to ensure the system's stability\footnote{Stability in the sense that a ``small" change in the operation conditions of the system leads to a ``small" change in its performance.}. The solution which stems from our general framework, described in detail in the sequel, is asymptotically (in the observation length) optimal for any SNR (not necessarily ``high"). The main contributions of this paper are summarized as follows:
\begin{itemize}
	\item ML estimation of the model parameters: We derive the likelihood equations of the unknown (deterministic) mixing matrix and the noises' variances, which are allowed to be different in each sensor in our model. We also provide the corresponding Cram\'er-Rao Lower Bound (CRLB) and all the required closed-form expressions for the Fisher scoring algorithm (e.g., \cite{jennrich1976newton}), which we propose as an iterative solution for the aforementioned likelihood equations.
	\item ML-based MMSE sources' estimation: Based on the MLEs mentioned above, we propose the ML-based MMSE estimate of the sources, which is identical to the classical MMSE estimate in its structure, but is based on the MLEs of the unknown (deterministic) model parameters rather than on their (unavailable) true values. We show, both analytically and numerically, that this estimate is asymptotically optimal, i.e., approaches the (oracle) MMSE lower bound as the observation length increases, in any SNR conditions.
	\item Efficient computation of the ML-based MMSE: Exploiting the stationarity of the signals, we provide a computationally efficient scheme, to be conveniently applied in the frequency domain.
	\item Quasi ML (QML)-based Linear MMSE (LMMSE) estimation: We show that the proposed scheme can also be successfully applied to non-Gaussian sources, when only their SOS (namely, their spectra) are known. In these scenarios, the model parameters estimates are essentially the Gaussian QML Estimates (QMLEs) \cite{weiss2018onconsistency,pham1997blind2}, yielding the sources' QML-based LMMSE estimates. These are shown to be slightly sub-optimal pseudo-linear\footnote{By ``pseudo-linear" we refer to linear estimates of the sources, in which the unknown fixed parameters are replaced with estimates thereof.} estimates of the sources, with an MSE converging to the (oracle) LMMSE bound. Furthermore, we demonstrate empirically in a simulated realistic VLC-MIMO system that the QML-based LMMSE is competitive (in terms of BER) with other pseudo-LMMSE estimates, based on the classical JADE and SOBI algorithms.
\end{itemize}

The rest of this paper is organized as follows. The remainder of this section is devoted to a brief outline of our notations. In Section \ref{sec:problemformulation} we present the semi-blind, Gaussian, noisy ICA problem formulation and an (approximately) equivalent frequency-domain formulation. The likelihood-equations are derived in Section \ref{sec:MLEderivation}, followed by the presentation of the corresponding CRLB and an iterative solution algorithm in subsections \ref{subsec:CRLB} and \ref{subsec:FisherScoring}, respectively. The (Q)ML-based (L)MMSE estimate is presented in Section \ref{sec:MLbasedMMSE} along with its asymptotic (sub-)optimality and a qualitative analysis of its resulting MSE. Then, an efficient computation scheme thereof is given in subsection \ref{subsec:efficientcomputation}. Simulations results, supporting our analytical results, are presented in Section \ref{sec:simulationresults}, and Section \ref{sec:conclusion} concludes the paper with final remarks.
\vspace{-0.4cm}
{\subsection{Notations and Preliminaries}
We use $a, \ua$ and $\A$ for a scalar, column vector and matrix, respectively, where $A_{ij}$ denotes the $(i,j)$-th element of the matrix $\A$ and $a[i]$ denotes the $i$-th element of the vector $\ua$. A hat symbol $\;\widehat{}\;$ denotes an estimate, thus, for example, $\widehat{a}$ denotes an estimate of $a$. The gradient of a scalar function $f\left(\ua\right)$ with respect to (w.r.t.) its vector argument $\ua$ is denoted by $\nabla_{\bm{a}}f$. The superscripts $(\cdot)^{\tps}$, $(\cdot)^{\dagger}$ and $(\cdot)^{-1}$ denote the transposition, Hermitian transposition and inverse operators, respectively. The notations $\Eset[\cdot], \Tr(\cdot), \det(\cdot)$ and $\Re\{\cdot\}$ denote the expectation, trace, determinant and real-part of their arguments, respectively. The Kronecker and Hadamard products are denoted by $\otimes$ and $\odot$, respectively. We also denote by $\I_{K}$ the $K\times K$ identity matrix, and the pinning vector $\ue_k$ denotes the $k$-th column of $\I_{K}$. Using these notations, we define $\E_{ij}\triangleq\ue_i\ue_j^{\tps}$ and $\delta_{ij}\triangleq\ue_i^{\tps}\ue_j$. We also define $\text{vec}(\cdot)$ as the operator which concatenates the columns of an $M \times N$ matrix into an $MN \times 1$ column vector. Furthermore, we define the operator $\text{Diag}(\cdot)$, which creates an $N\times N$ diagonal matrix from its $N$-dimensional vector argument. Finally, $\uo_N\in\Rset^{N\times 1}$ denotes the all zeros-vector and $\textrm{O}$ denotes the all zeros-matrix (with context-dependent dimensions).
\vspace{-0.25cm}
\section{Problem Formulation}\label{sec:problemformulation}
Consider the following $M$ sources - $L$ sensors static, instantaneous, linear model
\begin{equation}
\label{modelequation}
\X = \A\mS + \V \in \Rset^{L\times T},
\end{equation}
where $\mS=\left[\us_1\;\cdots\;\us_M\right]^{\rm{T}}\in\Rset^{M\times T}$ denotes a matrix of $M$ source signals of length $T$, $\A\in\Rset^{L\times M}$ is a (deterministic) mixing matrix, $\V=\left[\uv_1\;\cdots\;\uv_L\right]^{\rm{T}}\in\Rset^{L\times T}$ denotes a matrix of $L$ additive noise signals (one for each sensor), where we assume $L\geq M$, and the observed mixture signals are given by $\X=\left[\ux_1\;\cdots\;\ux_L\right]^{\rm{T}}\in\Rset^{L\times T}$. In our semi-blind model, we assume that all the source signals are zero-mean stationary Gaussian processes with known Positive-Definite (PD) Toeplitz temporal covariance matrices $\C_s^{(m)}\triangleq\Eset\left[\us_m\us_m^{\tps}\right]$ (for every $m\in\{1,\ldots,M\}$), distinct from one another. As in the standard ICA model, the sources $\us_1,\ldots,\us_M\in\Rset^{T\times1}$ (i.e., the rows of $\mS$) are assumed to be mutually statistically independent and the mixing matrix $\A$ is assumed to be unknown. Furthermore, we assume that the noise $\uv_1,\ldots,\uv_L\in\Rset^{T\times1}$ from all the sensors (i.e., the rows of $\V$) are mutually statistically independent, temporally-white Gaussian noise processes, each with a temporal covariance matrix $\Eset\left[\uv_{\ell}\uv_{\ell}^{\tps}\right]=\sigma_{v_\ell}^2\I_T$ (for every $\ell\in\{1,\ldots,L\}$), and are also statistically independent from all the sources. The noises' variances $\sigma_{v_1}^2,\ldots,\sigma_{v_L}^2\in\Rset^{+}$ are assumed to be (deterministic) unknown.

Thus, given the measurement matrix $\X$ and the sources' covariances $\{\C_s^{(m)}\}_{m=1}^M$, our goal is to separate and estimate the unobservable sources $\us_1,\ldots,\us_M$. Note that for this model, in contrary to the classical (fully blind) model, no permutation nor scale ambiguities exist, and the only remaining inevitable ambiguities are sign ambiguities\footnote{We shall address this issue in Section \ref{empiricalvalidation}.}.

For convenience in the derivations, let us consider a different (yet equivalent) representation of the problem. Using the Discrete-Fourier Transform (DFT), a frequency-domain representation of \eqref{modelequation} may be obtained (see also \cite{weiss2018boundsonpassive}). More precisely, denoting $\F\in\Cset^{T\times T}$ the $T$-dimensional (normalized) DFT matrix with elements $F_{kt}=\frac{1}{\sqrt{T}}\exp{\big(-j2\pi(k-1)(t-1)/T\big)}$, we define
\begin{equation}
\label{DFTequaltoFT}
\utx_{\ell}\triangleq\F\ux_{\ell}, \; \forall\ell\in\{1,\ldots,L\}.
\end{equation}
Since the DFT is merely a linear (complex-valued) invertible transformation, the available data may be written in the frequency domain as
\begin{equation}
\label{modelequationfreq}
\tX = \A\tS + \tV\in\Cset^{L\times T},
\end{equation}
or, equivalently, as
\begin{equation}
\label{modelequationfreqmatrixformdiscrete}
\utx[k] = \A\uts[k]+\utv[k]\in \Cset^{L\times 1}, \forall k\in\left\{1,\ldots,T\right\},
\end{equation}
where $\utz[k]$ denotes the $k$-th column (corresponding to the $k$-th frequency component) of a matrix $\tZ$ with $T$ (frequency components) columns\footnote{From now on we shall assume for convenience that $T$ is even.}. Now, recall that the sources, as well as the noises, are all stationary Gaussian and statistically independent. Hence, all the mixtures are jointly stationary and jointly Gaussian as well. In addition, since the DFT is a linear complex-valued transformation, the $k$-th frequency source vector $\uts[k]$ and noise vector $\utv[k]$ are (circular) Complex Normal (CN) (except for $k=1,\tfrac{T+2}{2}$, for which they are real-valued Normal),
\begin{gather}
\label{CN_sources}
\uts[k]\sim \mathcal{CN}\left(\uo_M,\P_k^s\right), \forall k\in\{1,\ldots,T\}\backslash\left\{1,\tfrac{T+2}{2}\right\},\\
\utv[k]\sim \mathcal{CN}\left(\uo_L,\mLambda\right), \forall k\in\{1,\ldots,T\}\backslash\left\{1,\tfrac{T+2}{2}\right\},
\end{gather}
where $\mLambda\triangleq \text{Diag}\left(\ulambda\right)\in\Rset^{L\times L}$, $\ulambda^{\tps}\triangleq\left[\sigma_{v_1}^2 \cdots \sigma_{v_L}^2\right]\in\Rset^{1 \times L}$ and asymptotically (namely, for $T$ large enough, due to the stationarity of the sources) $\P_k^s\in\Rset^{M\times M}$ are approximately\footnote{The approximation becomes arbitrarily close when $T$ is sufficiently large.} diagonal matrices containing the power spectral densities of the sources at the $k$-th frequency, i.e., their $(m,m)$-th element equals $P^s_m[k]$, where 
\begin{gather}
	P^s_m[k]\triangleq C_{s_{11}}^{(m)}+\lim\limits_{T \rightarrow\infty} 2\sum_{t=2}^{T}{C_{s_{1t}}^{(m)}\cos\left(\frac{2\pi (k-1)(t-1)}{T}\right)},\nonumber\\
	\forall k\in\left\{1,\ldots,T\right\}, \; \forall m\in\left\{1,\ldots,M\right\}.
\end{gather}
Consequently, since $\A$ is also a linear transformation and using the statistical independence between the sources and the noises, from \eqref{modelequationfreqmatrixformdiscrete} we have that
\begin{equation}
\label{CN_mixtures}
\utx[k]\sim \mathcal{CN}\left(\uo_L,\C_k\left(\A,\ulambda\right)\right), \forall k\in\{1,\ldots,T\}\backslash\left\{1,\tfrac{T+2}{2}\right\},
\end{equation}
where $\C_k\left(\A,\ulambda\right) \triangleq \A\P_k^s\A^{\tps}+\mLambda$. Notice that since all the considered time-domain signals are real-valued, the sufficient statistics are actually the first $T/2+1$ frequency components (the other $T/2-1$ are the complex conjugates of $\left\{\utx[k]\right\}_{k=2}^{T/2}$). Note also, that due to the stationarity of the sources and noises, and combined with their Gaussianity, these frequency components are (asymptotically) mutually statistically independent.

The problem at hand can now be formulated compactly in the frequency domain as follows. Given the statistically independent measurements $\left\{\utx[k]\right\}_{k=1}^{T/2+1}$ whose distributions are prescribed by \eqref{CN_mixtures}, separate and/or estimate the sources.

At this point, unlike in the noiseless case, it is crucial to explicitly define what the desired objective is. One option is to estimate the sources ``as closely as possible", e.g., in the sense of MMSE. Another possible (and simpler) objective is to obtain ``maximal separation" of the sources, even at the cost of a compromised MSE in their estimates. This approach is often termed in the context of communication systems as ``zero-forcing" (e.g., \cite{klein1996zero,wiesel2008zero}) and is known to minimize all (spatial) intersymbol interference. In ICA, a ``maximally separating" solution minimizes the resulting ISR, and in semi-blind scenarios this solution is obtained by applying the (pseudo-) inverse of the MLE of the mixing matrix to the mixtures' matrix $\X$ (as shown in, e.g., \cite{yeredor2010blind}).

Although both approaches yield two optimal estimates of the sources, we stress that they serve two fundamentally different objectives and accordingly result in two fundamentally different solutions. For example, the optimal estimate of the sources in the sense of MMSE applies filtering to the received signals, which is certainly not an instantaneous (memoryless) operation, like the mixing is, and distorts the signals with frequency-selective filtering and separation. In contrast, the optimal estimate of the sources in the sense of minimum ISR is an instantaneous (memoryless) operation, exactly like the mixing is, but may suffer from residual noise enhancement in frequencies where the sources have low magnitude (which may severely affect the output SNR). We note that if $L>M$, a ``maximally separating" memoryless solution is not unique, and may be chosen so as to (asymptotically) attain the minimum MSE among all maximally separating memoryless solutions. However, we shall not pursue this option in here, as it deviates from the main focus of this work.

Our main focus in this work is on presenting a separation-estimation scheme which aims to achieve (asymptotically) optimal estimation of the sources in the sense of MMSE, thus prioritizing proximity of the estimates to the true sources over maximal separation. However, due to the structure of our proposed solution, we can also obtain a ``maximally separating" memoryless solution based on the MLEs of the unknown model parameters. Therefore, regardless of the objective, our first step would be to derive the MLEs of these unknown (deterministic) parameters: $\A$ and $\ulambda$.
\vspace{-0.2cm}
\section{ML Estimation of the Model Parameters}\label{sec:MLEderivation}
In order to simplify the exposition, let $\utheta\in\Rset^{K_{\theta}\times 1}$ be the vector of unknown parameters $\A$ and $\ulambda$, where $K_{\theta}\triangleq ML+L$. More precisely,
\begin{equation}
\label{thetadefinition}
\utheta^{\tps}\triangleq \left[\text{vec}\left(\A\right)^{\tps} \ulambda^{\tps}\right]\in\Rset^{1\times K_{\theta}}.
\end{equation}
Since $\left\{\utx[k]\right\}_{k=1}^{T/2+1}$ are (asymptotically) statistically independent, the log-likelihood of $\utheta$ is the sum of log-likelihoods of $\utheta$ of all the frequency components. To this end, define
\begin{gather}
\mathcal{L}_k\left(\utheta\right) \triangleq \alpha_k\cdot\Big(\log\det\C_k^{-1} - \Tr\left(\mChi[k]\C_k^{-1}\right)\Big),\nonumber\\
\forall k \in \{1,\ldots,\tfrac{T+2}{2}\}, \label{loglikelihoodkfreq}
\end{gather}
as the ``relevant" log-likelihood (i.e., constants w.r.t. $\utheta$ are omitted) of the $k$-th frequency component, where we have used $\C_k$ instead of $\C_k\left(\A,\ulambda\right)$ for shorthand and defined $\mChi[k]\triangleq\utx[k]\utx[k]^{\dagger}\in\Cset^{L\times L}$ and $\alpha_k\triangleq1-0.5\cdot\left(\delta_{1k}+\delta_{(T/2+1)k}\right)$. The log-likelihood is then given by
\begin{equation}
\mathcal{L}\left(\utheta\right) \triangleq \sum_{k=1}^{T/2+1}{\mathcal{L}_k}\left(\utheta\right).
\end{equation}
As the global maximizer of $\mathcal{L}\left(\utheta\right)$, the MLE is a solution of $\nabla_{\bm{\theta}}\mathcal{L}^{\tps}=\uo_{K_{\theta}}$ (where $\nabla_{\bm{\theta}}\mathcal{L}$ is the score). Differentiating \eqref{loglikelihoodkfreq} w.r.t. $A_{ij}$, we have by the chain rule
\begin{equation}
\label{score_wrt_A}
\frac{\partial\mathcal{L}_k\left(\utheta\right)}{\partial A_{ij}} = \sum_{\ell_1,\ell_2=1}^{L}{\frac{\partial\mathcal{L}_k\left(\utheta\right)}{\partial C_{k_{\ell_1\ell_2}}}\cdot\frac{\partial C_{k_{\ell_1\ell_2}}}{\partial A_{ij}}},
\end{equation}
and based on well-known matrix functions derivatives (e.g., \cite{petersen2008matrix}), using 
\begin{align}
&\frac{\partial C_{k_{\ell_1\ell_2}}}{\partial A_{ij}} =\ue_{\ell_1}^{\tps}\left(\E_{ij}\P_k^s\A^{\tps}+\A\P_k^s\E_{ji}\right)\ue_{\ell_2}\nonumber\\
&\quad\quad\quad \:\:=P^s_j[k]\left(A_{\ell_1j}\delta_{i\ell_2}+A_{\ell_2j}\delta_{i\ell_1}\right),\label{dCdaij}\\
&\frac{\partial \log\det\C_k}{\partial C_{k_{\ell_1\ell_2}}} = \left(\C_k^{-1}\right)_{\ell_1\ell_2},
\end{align}
\begin{align}
&\frac{\partial \Tr\left(\mChi[k]\C_k^{-1}\right)}{\partial C_{k_{\ell_1\ell_2}}} = -\ue_{\ell_1}^{\tps}\left(\C_k^{-1}\mChi[k]\C_k^{-1}\right)\ue_{\ell_2},
\end{align}
we obtain after simplification
\vspace{-0.15cm}
\begin{align}
&\frac{\partial\mathcal{L}_k\left(\utheta\right)}{\partial A_{ij}} =\alpha_kP_j^s[k]\cdot\nonumber\\
&\sum_{\ell=1}^{L}{A_{\ell j}\Big[\Tr\left(\C_k^{-1}\mChi[k]\C_k^{-1}\left(\E_{i\ell}+\E_{\ell i}\right)\right)-2\left(\C_k^{-1}\right)_{i\ell}\Big]}=\nonumber\\
&2\alpha_kP_j^s[k]\sum_{\ell=1}^{L}{A_{\ell j}\Big[\uxi_{k,\ell}^{\tps}\Re\left\{\mChi[k]\right\}\uxi_{k,i}-\left(\C_k^{-1}\right)_{i\ell}\Big]},
\end{align}
where $\uxi_{k,\ell}$ denotes the $\ell$-th column (and row) of $\C_k^{-1}$. Thus, the score w.r.t. the $(i,j)$-th element of the mixing matrix is
\begin{align}
\label{dLdA}
&\frac{\partial\mathcal{L}\left(\utheta\right)}{\partial A_{ij}}=\nonumber\\
&2\hspace{-0.15cm}\sum_{k=1}^{T/2+1}\hspace{-0.1cm}\sum_{\ell=1}^{L}{\alpha_kP_j^s[k]A_{\ell j}\Big[\uxi_{k,\ell}^{\tps}\Re\left\{\mChi[k]\right\}\uxi_{k,i}-\left(\C_k^{-1}\right)_{i\ell}\Big]}.
\end{align}
Likewise, using
\begin{equation}
\label{dCdsigmav}
\frac{\partial \C_k}{\partial \sigma_{v_\ell}^2} = \E_{\ell \ell} \Rightarrow \frac{\partial C_{k_{\ell_1\ell_2}}}{\partial \sigma_{v_\ell}^2} = \delta_{\ell_1\ell_2}\delta_{\ell_1\ell},
\end{equation}
we obtain the score w.r.t. the noises' variances, given by
\begin{align}
\label{dLdsigmav}
\frac{\partial\mathcal{L}\left(\utheta\right)}{\partial \sigma_{v_\ell}^2} &= \sum_{k=1}^{T/2+1}{\alpha_k\Big[\ue_{\ell}^{\tps}\C_k^{-1}\mChi[k]\C_k^{-1}\ue_{\ell}-\left(\C_k^{-1}\right)_{\ell\ell}\Big]}\nonumber\\
&=\sum_{k=1}^{T/2+1}{\alpha_k\Big[\uxi_{k,\ell}^{\tps}\mChi[k]\uxi_{k,\ell}-\left(\C_k^{-1}\right)_{\ell\ell}\Big]}.
\end{align}
Therefore, the MLEs of $\A$ and $\ulambda$ are the solutions of the following system of (likelihood-) equations
\tcbset{colframe=black!100!blue,size=small,width=0.5\textwidth,halign=flush center,arc=3mm,outer arc=1mm}
\begin{tcolorbox}[upperbox=visible,colback=white]
\vspace{-0.2cm}
\begin{gather}
\forall i\in\{1,\ldots,L\}, \forall j\in\{1,\ldots,M\}:\nonumber\\
\hspace*{-0.1cm}\sum_{k=1}^{T/2+1}{\sum_{\ell=1}^{L}{\alpha_kP_j^s[k]\widehat{A}_{\ell j}\Big[\huxi_{k,\ell}^{\tps}\Re\left\{\mChi[k]\right\}\huxi_{k,i}- \left(\hC_k^{-1}\right)_{i\ell}\Big]}}=0,\nonumber\\
\sum_{k=1}^{T/2+1}{\alpha_k\Big[\huxi_{k,i}^{\tps}\mChi[k]\huxi_{k,i}-\left(\hC_k^{-1}\right)_{ii}\Big]}=0,\label{likelihoodequations}
\end{gather}
\end{tcolorbox}
\noindent which bring $\mathcal{L}\left(\utheta\right)$ to its global maximum, where $\hC_k$, and accordingly $\hC_k^{-1}$ (and each $\huxi_{k,\ell}$), encapsulate $\hA$ and $\hulambda$, which denote the MLEs of $\A$ and $\ulambda$, respectively. As known, these estimates are asymptotically efficient, thus asymptotically achieving the CRLB on the MSE, presented (implicitly) in what follows.
\vspace{-0.25cm}
\subsection{The CRLB on the MSE of the Model Parameters' Estimates}\label{subsec:CRLB}
Since $\left\{\utx[k]\right\}_{k=1}^{T/2+1}$ are all CN (except for $k=1,\tfrac{T+2}{2}$, for which they are real-valued Normal), the Fisher Information Matrix (FIM) elements corresponding to $\A$ and $\ulambda$ are given by (see, e.g., \cite{collier2005fisher})
\begin{gather}
\mathcal{I}[A_{ij},A_{pq}]=\hspace{-0.05cm}\sum_{k=1}^{T/2+1}{\hspace{-0.1cm}\alpha_k\cdot\Tr\left(\C_k^{-1}\frac{\partial\C_k}{\partial A_{ij}}\C_k^{-1}\frac{\partial\C_k}{\partial A_{pq}}\right)},\label{FIM_CN_element1}\\
\mathcal{I}[A_{ij},\sigma_{v_\ell}^2]=\hspace{-0.05cm}\sum_{k=1}^{T/2+1}{\hspace{-0.1cm}\alpha_k\cdot\Tr\left(\C_k^{-1}\frac{\partial\C_k}{\partial A_{ij}}\C_k^{-1}\frac{\partial\C_k}{\partial \sigma_{v_\ell}^2}\right)},\label{FIM_CN_element2}\\
\mathcal{I}[\sigma_{v_{\ell_1}}^2,\sigma_{v_{\ell_2}}^2]=\hspace{-0.05cm}\sum_{k=1}^{T/2+1}{\hspace{-0.1cm}\alpha_k\cdot\Tr\left(\C_k^{-1}\frac{\partial\C_k}{\partial \sigma_{v_{\ell_1}}^2}\C_k^{-1}\frac{\partial\C_k}{\partial \sigma_{v_{\ell_2}}^2}\right)},\label{FIM_CN_element3}
\end{gather}
where $\bs{\mathcal{I}}(\utheta)$ denotes the FIM. Notice that the closed form expressions for the terms $\frac{\partial \mathbfit{C}_k}{\partial A_{ij}}$ and $\frac{\partial\mathbfit{C}_k}{\partial \sigma_{v_\ell}^2}$ were already obtained in the previous subsection (see \eqref{dCdaij} and \eqref{dCdsigmav}). In addition, by the Woodbury matrix identity \cite{woodbury1950inverting}, we have 
\begin{equation}
\label{FIM_term_c}
\C_k^{-1}=\mLambda^{-1}-\mLambda^{-1}\A\left(\P_k^{s^{-1}}+\A^{\tps}\mLambda^{-1}\A\right)^{-1}\A^{\tps}\mLambda^{-1}.
\end{equation}
Therefore, all the required expressions for the computation of the FIM (for any known values of the mixing matrix and the noise variances) are at hand. For example, by \eqref{dCdsigmav}, we have
\begin{equation}
\mathcal{I}[\sigma_{v_{\ell_1}}^2,\sigma_{v_{\ell_2}}^2] = \sum_{k=1}^{T/2+1}{\alpha_k\cdot\left(\left(\hC_k^{-1}\right)_{\ell_1\ell_2}\right)^2}.
\end{equation}
The CRLB on the MSE in unbiased joint estimation of $\A$ and $\ulambda$ is given by the inverse of the FIM, whose elements are prescribed in \eqref{FIM_CN_element1}-\eqref{FIM_CN_element3}. Next, we consider an approximate iterative solution algorithm based on the results obtained in this subsection.
\subsection{MLE Computation via the Fisher Scoring Algorithm}\label{subsec:FisherScoring}
From \eqref{dLdA} and \eqref{dLdsigmav} we have the closed form expressions of the score w.r.t. each of the elements of $\utheta$. Moreover, we have obtained closed form expressions for the elements of the FIM \eqref{FIM_CN_element1}-\eqref{FIM_CN_element3}, which can be computed for any $\utheta$. Therefore, given an initial estimate of $\utheta$, the Fisher scoring algorithm may be used in order to obtain a stationary point of the log-likelihood (if it converges). If the initial solution is some ``educated" guess which is ``close" enough to the global maximizer of $\mathcal{L}\left(\utheta\right)$, the algorithm is likely to converge to the MLE. The update equation of the Fisher scoring algorithm for the $n$-th iteration is given by
\begin{equation}\label{FSAupdatequation}
\hutheta^{(n)} = \hutheta^{(n-1)} + \bs{\mathcal{I}}^{-1}\left(\hutheta^{(n-1)}\right)\left.\nabla_{\bm{\theta}}\mathcal{L}^{\tps}\right|_{\scriptsize{\utheta}=\hutheta^{(n-1)}},
\end{equation}
where $\hutheta^{(n)}$ denotes the estimate of $\utheta$ in the $n$-th iteration. We note in passing that as long as $M\cdot L$ is not too ``large" (in terms of matrix inversion), the computation of $\bs{\mathcal{I}}^{-1}\left(\hutheta^{(n)}\right)$ is not very costly, w.r.t. computational load.

Having derived the likelihood-equations, the CRLB, and an iterative solution algorithm for the MLEs of the mixing matrix and the noises' variances, we now turn to our primary task - separation and estimation of the unobserved sources. We shall pursue the (asymptotically) optimal estimate, in the sense of minimal MSE, based on the MLEs of $\A$ and $\ulambda$. 
\section{Sources ML-Based (L)MMSE Estimation}\label{sec:MLbasedMMSE}
Assume for the moment that $\A$ and $\ulambda$ are known, and consider the following equivalent representation of \eqref{modelequation}
\begin{equation}
\ux = \left(\A\otimes\I_T\right)\us + \uv\in\Rset^{LT\times1},
\end{equation}
where $\ux\triangleq\text{vec}\left(\X^{\tps}\right)\in\Rset^{LT\times1}$, $\uv\triangleq\text{vec}\left(\V^{\tps}\right)\in\Rset^{LT\times1}$ and $\us\triangleq\text{vec}\left(\mS^{\tps}\right)\in\Rset^{MT\times1}$. In this case, since $\ux$ and $\us$ are jointly Gaussian, the MMSE estimate of $\us$ from $\ux$ (which is also the LMMSE estimate of $\us$ from $\ux$) is given by
\begin{equation}
\label{MMSEestimate}
\widehat{\us}_{\text{\tiny{MMSE}}}\triangleq\Eset\left[\us|\ux\right]=\C_{\scriptsize{\us\ux}}\C_{\scriptsize{\ux}}^{-1}\ux\in\Rset^{MT\times1},
\end{equation}
where
\begin{align}
&\C_{\scriptsize{\us\ux}}\triangleq\Eset\left[\us\ux^{\tps}\right]=\C_{\scriptsize{\us}}\left(\A^{\tps}\otimes\I_T\right)\in\Rset^{MT\times LT},\\
&\C_{\scriptsize{\ux}}\triangleq\Eset\left[\ux\ux^{\tps}\right]=\nonumber\\
&\quad\;\:\,\left(\A\otimes\I_T\right)\C_{\scriptsize{\us}}\left(\A^{\tps}\otimes\I_T\right)+\mLambda\otimes\I_{T}\in\Rset^{LT\times LT},
\end{align}
\begin{equation}
\C_{\scriptsize{\us}}\triangleq\Eset\left[\us\us^{\tps}\right]=\begin{bmatrix}
\C_s^{(1)} & \textrm{O} & \dots  & \textrm{O} \\
\textrm{O} & \C_s^{(2)} & \dots  & \textrm{O} \\
\vdots & \vdots & \ddots & \vdots \\
\textrm{O} & \textrm{O} & \dots  & \C_s^{(M)}
\end{bmatrix}\in\Rset^{MT\times MT}.
\end{equation}
Accordingly, the $m$-th source MMSE estimate is given by
\begin{equation}
\left(\widehat{\us}_m\right)_{\text{\tiny{MMSE}}}=\hS^{\tps}_{\text{\tiny{MMSE}}}\ue_{m}\in\Rset^{T\times 1},
\end{equation}
where $\widehat{\us}_{\text{\tiny{MMSE}}}=\text{vec}\left(\hS^{\tps}_{\text{\tiny{MMSE}}}\right)$. This estimate attains the minimal attainable MSE matrix, given by
\begin{equation}
\label{MSE_matrix}
\C_{\scriptsize{\uepsilon}}\triangleq\Eset\left[\left(\widehat{\us}_{\text{{\tiny{MMSE}}}}-\us\right)\left(\widehat{\us}_{\text{{\tiny{MMSE}}}}-\us\right)^{\tps}\right]=\C_{{\scriptsize{\us}}}-\C_{{\scriptsize{\us\ux}}}\C_{{\scriptsize{\ux}}}^{-1}\C_{{\scriptsize{\ux\us}}},
\end{equation}
which we refer to as the (oracle) MMSE bound for the Gaussian case, and which is also the (oracle) LMMSE bound in the general case. Note that, as a quick ``sanity check", it is easily seen that when $\mLambda=\textrm{O}$ and $\A$ is square invertible, we get $\hat{\us}=(\A^{-1}\otimes\I_T)\ux$ (perfect separation) with $\C_{\scriptsize{\uepsilon}} = \textrm{O}$, as expected (when $\A$ and $\ulambda$ are known).
\subsection{The ML-Based MMSE Solution}\label{subsec:MLbasedLMMSE}
Now, recall that $\A$ and $\ulambda$ are in fact unknown. Therefore, we suggest the following ML-based MMSE estimate of the sources
\begin{equation}
\label{MLbasedMMSE}
\widehat{\us}_{\text{\tiny{ML-MMSE}}}\triangleq \hC_{\scriptsize{\us\ux}}\hC_{\scriptsize{\ux}}^{-1}\ux,
\end{equation}
based only on the measurements $\ux$, where
\begin{gather}
\hC_{\scriptsize{\us\ux}}\triangleq\C_{\scriptsize{\us}}\left(\hA_{\text{\tiny{ML}}}^{\tps}\otimes\I_T\right),\label{covest1}\\
\hC_{\scriptsize{\ux}}\triangleq\left(\hA_{\text{\tiny{ML}}}\otimes\I_T\right)\C_{\scriptsize{\us}}\left(\hA_{\text{\tiny{ML}}}^{\tps}\otimes\I_T\right)+\widehat{\mLambda}_{\text{\tiny{ML}}}\otimes\I_T \label{covest2},
\end{gather}
and where $\hA_{\text{\tiny{ML}}}$ and $\widehat{\mLambda}_{\text{\tiny{ML}}}$ are the MLEs of $\A$ and $\mLambda$, respectively. We stress that for any finite sample size $T$, $\widehat{\us}_{\text{\tiny{ML-MMSE}}}\neq\widehat{\us}_{\text{\tiny{MMSE}}}$ almost surely. However, the estimate \eqref{MLbasedMMSE} enjoys an (attractive) asymptotic optimality property, as we show in the sequel (subsection \ref{subsec:asymptoticoptimality}). 

Next, we present an efficient computation of the ML-based MMSE estimate of the sources, given the MLEs $\hA_{\text{\tiny{ML}}}$ and $\widehat{\mLambda}_{\text{\tiny{ML}}}$, based on the stationarity of the signals.
\subsection{Efficient Computation of the ML-Based MMSE Estimate in the Frequency Domain}\label{subsec:efficientcomputation}
Denote $\F_N\triangleq\I_N\otimes\F\in\Cset^{NT\times NT}$ (for any $N\in\Zset$), and notice that $\F_N\F_N^{\dagger}=\F_N^{\dagger}\F_N=\I_{NT}$. Now, applying $\F_M$ to \eqref{MMSEestimate} from the left, we have
\begin{gather}
\widehat{\widetilde{\us}}_{\text{\tiny{MMSE}}}\triangleq\F_M\widehat{\us}_{\text{\tiny{MMSE}}}=\F_M\C_{\scriptsize{\us\ux}}\C_{\scriptsize{\ux}}^{-1}\ux=\nonumber\\
\left(\F_M\C_{\scriptsize{\us\ux}}\F_L^{\dagger}\right)\left(\F_L\C_{\scriptsize{\ux}}^{-1}\F_L^{\dagger}\right)\left(\F_L\ux\right)= \tC_{\scriptsize{\us\ux}}\tC_{\scriptsize{\ux}}^{-1}\widetilde{\ux},
\end{gather}
where
\begin{gather}
\label{covarianceestimatesFreqdomain}
\tC_{\scriptsize{\us\ux}}\triangleq\tC_{\scriptsize{\us}}\left(\A^{\tps}\otimes\I_T\right),\\
\tC_{\scriptsize{\ux}}\triangleq\left(\A\otimes\I_T\right)\tC_{\scriptsize{\us}}\left(\A^{\tps}\otimes\I_T\right)+\mLambda\otimes\I_{T},\\
\tC_{\scriptsize{\us}}\triangleq\begin{bmatrix}
\tC_s^{(1)} & \textrm{O} & \dots  & \textrm{O} \\
\textrm{O} & \tC_s^{(2)} & \dots  & \textrm{O} \\
\vdots & \vdots & \ddots & \vdots \\
\textrm{O} & \textrm{O} & \dots  & \tC_s^{(M)}
\end{bmatrix}\in\Rset^{MT\times MT},
\end{gather}
and $\{\tC_s^{(m)}=\text{Diag}\left(\left[P^s_m[1],\ldots,P^s_m[T]\right]\right)\in\Rset^{T\times T}\}_{m=1}^M$ are all PD diagonal matrices. Define the block matrices of the inverse of $\tC_{\scriptsize{\ux}}$ as follows
\begin{equation}
\label{invcovarianceestimatesFreqdomain}
\tC_{\scriptsize{\ux}}^{-1}\triangleq\begin{bmatrix}
\tP_x^{(1,1)} & \dots  & \tP_x^{(1,L)} \\
\vdots & \ddots & \vdots \\
\tP_x^{(L,1)} & \dots  & \tP_x^{(L,L)}
\end{bmatrix}\in\Rset^{LT\times LT},
\end{equation}
where $\{\tP_x^{(\ell_1,\ell_2)}\in\Rset^{T\times T}\}_{\ell_1,\ell_2=1}^L$ are all diagonal matrices. In the same manner, denote\
\begin{gather}
\label{invcrosscovarianceestimatesFreqdomain}
\tC_{\scriptsize{\us\ux}}
\triangleq\begin{bmatrix}
\tC_{sx}^{(1,1)} & \dots  & \tC_{sx}^{(1,L)} \\
\vdots & \ddots & \vdots \\
\tC_{sx}^{(M,1)} & \dots  & \tC_{sx}^{(M,L)}
\end{bmatrix},
\end{gather}
such that $\tC_{sx}^{(m,\ell)}=A_{\ell m}\cdot\tC_s^{(m)}\in\Rset^{T\times T}$, according to \eqref{covarianceestimatesFreqdomain}. Thus, the ML-based MMSE estimate of the sources in the frequency domain is given by
\begin{gather}
\label{MLMMSEestimatefreq}
\widehat{\widetilde{\us}}_{\text{\tiny{ML-MMSE}}}=\htC_{\scriptsize{\us\ux}}\htC_{\scriptsize{\ux}}^{-1}\widetilde{\ux},
\end{gather}
and accordingly the ML-based MMSE frequency-domain estimate of the $m$-th source is given by
\begin{equation}
\label{MLMMSEestimatefreqmthsource}
\left(\widehat{\widetilde{\us}}_m\right)_{\text{\tiny{ML-MMSE}}}=\sum_{\ell_1,\ell_2=1}^{L}{\htC_{sx}^{(m,\ell_1)}\htP_{x}^{(\ell_1,\ell_2)}\widetilde{\ux}_{\ell_2}},
\end{equation}
where $\htC_{\scriptsize{\us\ux}},\htC_{\scriptsize{\ux}},\htC_{sx}^{(m,\ell)}$ and $\htP_{x}^{(\ell_1,\ell_2)}$ are the MLEs of $\tC_{\scriptsize{\us\ux}},\tC_{\scriptsize{\ux}},\tC_{sx}^{(m,\ell)}$ and $\tP_{x}^{(\ell_1,\ell_2)}$, respectively, based on the MLEs $\hA_{\text{\tiny{ML}}}$ and $\widehat{\mLambda}_{\text{\tiny{ML}}}$. 

Notice that the computation of the sources' estimates in the frequency domain may be implemented more efficiently than it may in the time domain, since (asymptotically) it involves multiplications of (vectors and) diagonal matrices only. Finally, the time-domain estimates of the sources may be computed (also efficiently, via the FFT algorithm \cite{cooley1965algorithm}) from the frequency-domain estimates. Note that the respective MSE of this estimate may be computed efficiently in the frequency domain as well by the same principles present above.
\vspace{-0.3cm}
\subsection{Asymptotic Optimality and MSE Analysis of the ML-Based MMSE Estimate}\label{subsec:asymptoticoptimality}
From the invariance property of the MLE \cite{mukhopadhyay2000probability}, it follows that $\hC_{\scriptsize{\us\ux}}$ and $\hC_{\scriptsize{\ux}}$ are the MLEs of $\C_{\scriptsize{\us\ux}}$ and $\C_{\scriptsize{\ux}}$, respectively. In particular, $\hC_{\scriptsize{\us\ux}}$ and $\hC_{\scriptsize{\ux}}$ are consistent estimates (\cite{cramer2016mathematical}) of $\C_{\scriptsize{\us\ux}}$ and $\C_{\scriptsize{\ux}}$, respectively. Therefore, from the continuous mapping theorem \cite{mann1943stochastic}, which states that continuous functions are limit-preserving even if their arguments are sequences of random variables, we have that
\begin{equation}
\left(\hA_{\text{\tiny{ML}}},\widehat{\mLambda}_{\text{\tiny{ML}}}\right) \xrightarrow[T\rightarrow\infty]{p}\left(\A,\mLambda\right)\Rightarrow
\end{equation}
\vspace{-0.3cm}
\begin{equation}
\hC_{\scriptsize{\us\ux}} \xrightarrow[T\rightarrow\infty]{p}\C_{\scriptsize{\us\ux}},\;\;\hC_{\scriptsize{\ux}} \xrightarrow[T\rightarrow\infty]{p}\C_{\scriptsize{\ux}}\Rightarrow
\end{equation}
\vspace{-0.3cm}
\begin{equation}
\label{MLMMSEapproachesMMSE}
	\widehat{\us}_{\text{\tiny{ML-MMSE}}} \xrightarrow[T\rightarrow\infty]{p}\widehat{\us}_{\text{\tiny{MMSE}}}\:,
\end{equation}
since the LMMSE is a continuous function of $\A$ and $\mLambda$. Hence, the estimate \eqref{MLbasedMMSE} asymptotically attains the minimal attainable MSE. Indeed, as $T$ grows, the MSE attained by the estimate \eqref{MLbasedMMSE} decreases and converges (in probability) to the MMSE. Nevertheless, note that the MMSE is strictly positive as long as there are at least $L-M+1$ strictly positive elements of $\ulambda$ (i.e., the number of noiseless measurements is strictly smaller than the number of sources), even for an infinitely large sample size $T$. In what follows, we try to intuitively explain the nature of this problem w.r.t. the bounds on the estimation errors by presenting a (simplified) geometrical interpretation.

Define $\uepsilon_{\text{\tiny{MMSE}}}\triangleq\widehat{\us}_{\text{\tiny{MMSE}}}-\us$, $\uepsilon_{\text{\tiny{ML-MMSE}}}\triangleq\widehat{\us}_{\text{\tiny{ML-MMSE}}}-\us$ and $\ud \triangleq\widehat{\us}_{\text{\tiny{ML-MMSE}}}-\widehat{\us}_{\text{\tiny{MMSE}}}$. Using these notations, we have that
\begin{align}
&\Eset\left[\uepsilon_{\text{\tiny{ML-MMSE}}}\uepsilon^{\tps}_{\text{\tiny{ML-MMSE}}}\right]\hspace{-0.025cm}=\hspace{-0.025cm}\Eset\left[\uepsilon_{\text{\tiny{MMSE}}}\uepsilon^{\tps}_{\text{\tiny{MMSE}}}\right]\hspace{-0.025cm}+\hspace{-0.025cm}\Eset\left[\ud\ud^{\tps}\right]\hspace{-0.025cm}+\nonumber\\
&\Eset\left[\uepsilon_{\text{\tiny{MMSE}}}\ud^{\tps}\right]\hspace{-0.025cm}+\hspace{-0.025cm}\Eset\left[\ud\uepsilon_{\text{\tiny{MMSE}}}^{\tps}\right]\hspace{-0.025cm}=\hspace{-0.025cm}\Eset\left[\uepsilon_{\text{\tiny{MMSE}}}\uepsilon^{\tps}_{\text{\tiny{MMSE}}}\right]\hspace{-0.025cm}+\hspace{-0.025cm}\Eset\left[\ud\ud^{\tps}\right],
\end{align}
where the last transition is due to the well-known orthogonality\footnote{in the sense that the inner product between two random vectors is defined as their cross-covariance matrix} of the estimation error to any function of the measurements in MMSE estimation. Since both $\widehat{\us}_{\text{\tiny{ML-MMSE}}}$ and $\widehat{\us}_{\text{\tiny{MMSE}}}$ are clearly functions of the measurements (only), so is their difference $\ud$, which is therefore orthogonal to $\uepsilon_{\text{\tiny{MMSE}}}$. The result is a particular case of the Pythagorean theorem \cite{krall1986hilbert}.
\begin{figure}
	\centering
	\includegraphics[width=0.5\textwidth]{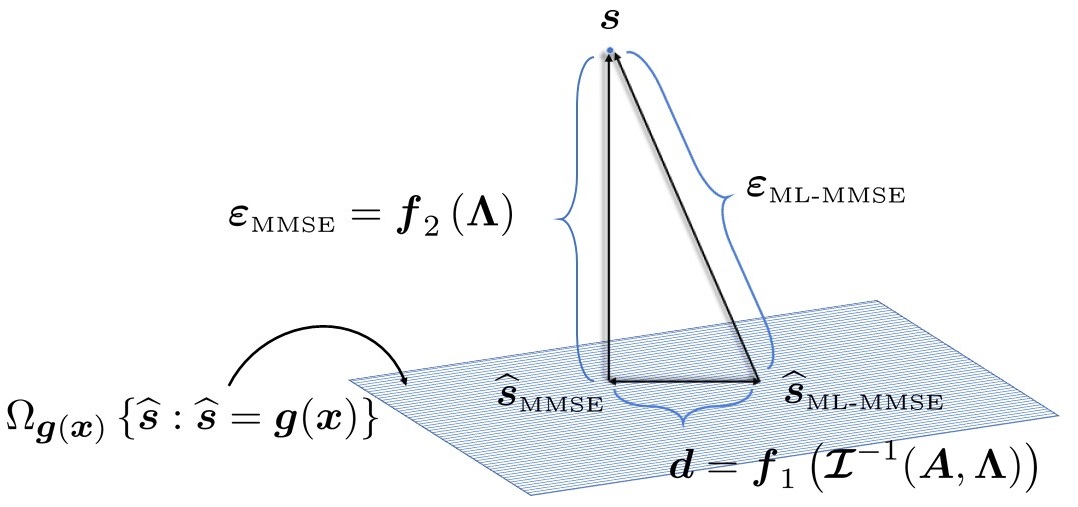}
	\caption{Simplified geometrical visualization of the optimality-gap between the estimation errors of the MMSE and the ML-based MMSE estimates. $\uf_1\left(\bs{\mathcal{I}}^{-1}(\A,\mLambda)\right)$ and $\uf_2\left(\mLambda\right)$ are monotonic increasing functions of $\bs{\mathcal{I}}^{-1}(\A,\mLambda)$ and $\mLambda$, respectively. We stress that this figure is solely for the purpose of intuition and is not meant to be quantitatively accurate.} 
	\label{fig:3Dvisualization}
	\vspace{-0.3cm}
\end{figure}
We term $\Eset\left[\ud\ud^{\tps}\right]$ the ``optimality-gap" matrix between the MSE of the ML-based MMSE estimate and the MMSE. Fig. \ref{fig:3Dvisualization} presents a (simplified) geometrical interpretation of this gap. Indeed, as we have shown, when the sample size $T$ approaches infinity, and $\mLambda$ is fixed (and finite), the angle of the upper vertex of the triangle presented in Fig. \ref{fig:3Dvisualization} approaches zero, and the triangle approaches a line orthogonal to the measurements space. The gap between the MMSE and ML-based MMSE estimates is a monotonically increasing function of the CRLB on the MSE in unbiased estimation of $\A$ and $\ulambda$. On the other hand, when at least $M$ elements of $\ulambda$ (the noises' variances) approach zero, and $T$ is fixed, the angle of the lower-right vertex of the triangle presented in Fig. \ref{fig:3Dvisualization} approaches zero, and the triangle approaches a line embedded in the measurements space. This is obviously expected, since in this limit $\us$ is merely a linear transformation of $\ux$. However, note that even when $\ulambda=\uo_L$ and $T\rightarrow\infty$ the MSE is not necessarily zero. Indeed, if, for example, $\A$ is not full rank, the mixing operation is not invertible, and separation of the sources in not achievable.

Fortunately, the proposed separation-estimation scheme yields a near-optimal solution to problem \eqref{modelequation} even in scenarios where the sources are \textit{not} Gaussian, as shown in the next subsection.
\vspace{-0.2cm}
\subsection{The QML-Based LMMSE Solution}\label{subsec:QMLbasedLMMSE}
Clearly, when the sources' SOS are known but the sources' distributions (whether they be known or unknown) are non-Gaussian, equations \eqref{likelihoodequations} become the \textit{quasi}-likelihood (rather than the likelihood) equations. Additionally, since the sources are not Gaussian, the LMMSE estimate is no longer guaranteed to be the MMSE estimate. Nevertheless, it is still the best linear estimate (in the sense of MMSE) of the sources based on the measurements $\X$, which is a reasonable (fallback) option for a problem with a linear model such as \eqref{modelequation}. Hence, in the case of non-Gaussian stationary sources with distinct spectra, the proposed scheme yields a QML-based LMMSE estimate of the sources. Nonetheless, this estimate is asymptotically the optimal linear estimate of the sources. To show this, we observe the following. By definition, we have $\Eset\left[\mChi[k]\right]=\C_k$ regardless of the sources' distributions, and since $\C_k\in\Rset^{L\times L}$ it follows that also $\Eset\left[\Re\left\{\mChi[k]\right\}\right]=\C_k$. Thus, define 
\begin{equation}\label{QMLcondition}
\Ep_{C_k}\triangleq\Re\left\{\mChi[k]\right\}-\C_k, \; \forall k\in \{1,\ldots,\tfrac{T+2}{2}\},
\end{equation}
such that $\left\{\Ep_{C_k}\right\}_{k=1}^{T/2+1}$ are all zero-mean matrices. Note also that the variances of all the elements of $\left\{\Ep_{C_k}\right\}_{k=1}^{T/2+1}$ are finite and independent of $T$. Now, substituting $\hA$ and $\hulambda$ with the true $\A$ and $\ulambda$, respectively, the left-hand sides of the quasi-likelihood equations \eqref{likelihoodequations} (i.e., for non-Gaussian sources) become
\begin{gather}
\forall i\in\{1,\ldots,L\}, \forall j\in\{1,\ldots,M\}:\nonumber\\
\sum_{k=1}^{T/2+1}{\sum_{\ell=1}^{L}{\alpha_kP_j^s[k]A_{\ell j}\Big[\uxi_{k,\ell}^{\tps}\Ep_{C_k}\uxi_{k,i}\Big]}}\triangleq\sum_{k=1}^{T/2+1}{\eta^{(i,j)}_1[k]},\nonumber\\
\sum_{k=1}^{T/2+1}{\alpha_k\Big[\uxi_{k,i}^{\tps}\Ep_{C_k}\uxi_{k,i}\Big]}\triangleq\sum_{k=1}^{T/2+1}{\eta^{(i)}_2[k]},\label{quasilikelihoodequations}
\end{gather}
where $\{\eta^{(i,j)}_1[k]\}_{k=1}^{T/2+1}$ are all zero-mean, uncorrelated random variables with finite (bounded) variances, as well as $\{\eta^{(i)}_2[k]\}_{k=1}^{T/2+1}$, for every $i\in\{1,\ldots,L\}$ and $j\in\{1,\ldots,M\}$. Therefore, by virtue of the (weak) law of large numbers \cite{ross2009first},
\begin{gather}
\forall i\in\{1,\ldots,L\}, \forall j\in\{1,\ldots,M\}:\nonumber\\
\sum_{k=1}^{T/2+1}{\eta^{(i,j)}_1[k]}\xrightarrow[T\rightarrow\infty]{p}0,\sum_{k=1}^{T/2+1}{\eta^{(i)}_2[k]}\xrightarrow[T\rightarrow\infty]{p}0.\label{asympquasilikelihoodequations}
\end{gather}
Thus, asymptotically, the score $\nabla_{\bm{\theta}}\mathcal{L}$ vanishes at $\hA=\A$ and $\hulambda=\ulambda$, so that the true mixing matrix and noise variances are indeed a solution for the asymptotic quasi-likelihood equations even when the sources are non-Gaussian. It follows immediately from the same arguments presented before \eqref{MLMMSEapproachesMMSE} that the QML-based LMMSE estimate, given by \eqref{MLbasedMMSE} with the QMLEs of $\A$ and $\mLambda$ replacing their MLEs, respectively, in \eqref{covest1}--\eqref{covest2}, converges (in probability) to the LMMSE estimate.

The QMLEs of $\A$ and $\mLambda$ may be obtained by the iterative algorithm presented in subsection \ref{subsec:FisherScoring}, only in this context it is no longer the Fisher scoring algorithm, since $\bs{\mathcal{I}}(\utheta)$, as defined in \eqref{FIM_CN_element1}--\eqref{FIM_CN_element3}, is no longer the FIM. Nevertheless, since $\bs{\mathcal{I}}(\utheta)$ is PD by definition for any vector-argument $\utheta\in\Rset^{K_{\theta}\times1}$ (with the last $L$ elements being non-negative), equation \eqref{FSAupdatequation} serves as a quasi-Newton algorithm (e.g., \cite{bonnans2006numerical}) update equation. Regarding the efficient computation of the proposed estimate, the complete derivation presented in subsection \ref{subsec:efficientcomputation} is totally valid for the QML-based LMMSE as well, when $\hA_{\text{\tiny{ML}}}$ and $\widehat{\mLambda}_{\text{\tiny{ML}}}$ are replaced by the QMLEs of $\A$ and $\mLambda$, respectively, since it relies only on the stationarity of the signals.

Note that despite what might be (wrongfully) implied from its name, the QML-based LMMSE estimate is certainly \text{not} a linear estimate of the sources, in addition to not being the actual LMMSE. This is because $\hA_{\text{\tiny{QML}}}$ and $\widehat{\mLambda}_{\text{\tiny{QML}}}$, the QMLEs of $\A$ and $\mLambda$, respectively, are nonlinear functions of $\X$, as they are solutions of the quasi-likelihood (nonlinear) equations \eqref{likelihoodequations}. Therefore, the QML-based LMMSE estimate, which is a function of $\hA_{\text{\tiny{QML}}}$ and $\widehat{\mLambda}_{\text{\tiny{QML}}}$, is also a nonlinear function of $\X$, and is therefore termed ``pseudo-linear" in here.

We note in passing that although the result above does not depend on the mixtures' DFTs distributions, for a wide range of stationary sources, the resulting mixtures' DFTs \eqref{modelequationfreqmatrixformdiscrete} will converge in distribution to the (circular) CN distribution, as prescribed in \eqref{CN_mixtures}, by virtue of the (Lyapunov's) central limit theorem \cite{koralov2007theory}. This, in turn, strengthens the quasi-likelihood approximation, since eventually it is based on \eqref{CN_mixtures}. However, we stress that despite the \textit{marginal} distribution convergence of each frequency component to its limiting CN distribution, the \textit{joint} distribution of all of these components is not the multivariate CN distribution, since they are not statistically independent. For this reason, although an increasing sample size does strengthen our quasi-likelihood approximation, it would still be, even asymptotically, merely a QMLE approximation.
\vspace{-0.3cm}
\section{Empirical Validation by Simulation Results}\label{sec:simulationresults}
\label{empiricalvalidation}
In this section we validate our theoretical derivations by empirical simulation results of three experiments. The first two experiments consider two similar scenarios, where in both the sources are Gaussian Auto-Regressive (AR) processes. However, in experiment 1 the high SNR regime is considered, while in experiment 2 the low SNR regime and a significantly smaller sample size than in experiment 1 are considered, demonstrating the proposed scheme's robustness to different SNR and sample size conditions. In both of these experiments, we present accuracy estimation measures for the two stages of the proposed scheme - ML estimation of the mixing matrix and the noise signals' variances and consequent MMSE separation-estimation of the sources. In the third experiment, we demonstrate how the proposed scheme can be applied in the context of VLC MIMO systems, an emerging application in the field of optical communication, for joint estimation of the channel and SNR, and for consequent (pseudo-) LMMSE-based estimation of transmitted sequences (bits). In this experiment, the overall performance are evaluated by the resulting BER of the estimated transmitted sequences.
\vspace{-0.3cm}
\subsection{Experiment 1: AR Sources in High SNR}
	\begin{figure}
	\centering
	\includegraphics[width=0.5\textwidth]{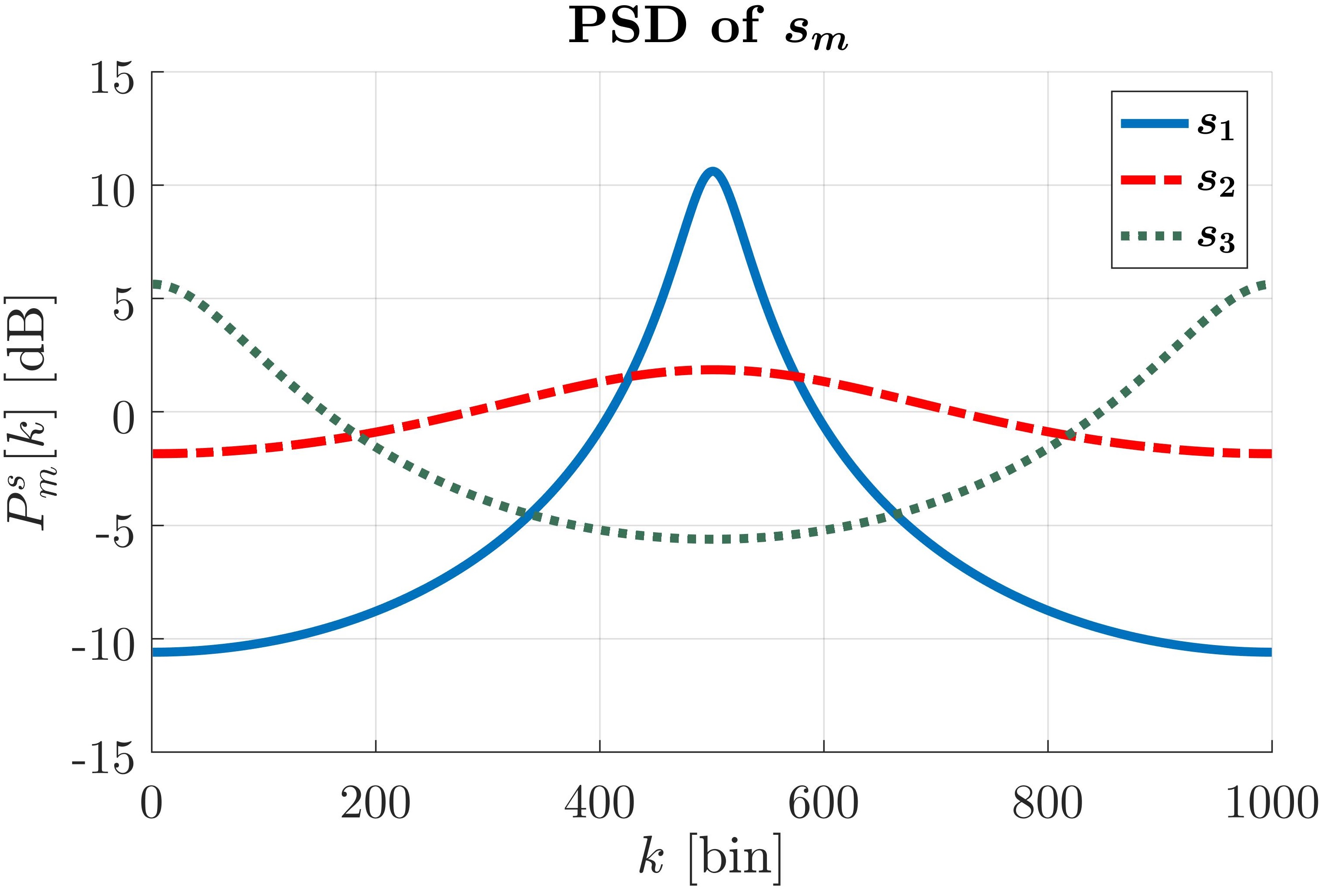}
	\caption{Spectra of the $M=3$ AR(1) sources for $T=1000$. The $k$-th bin corresponds to the $k$-th frequency component $\tfrac{2\pi(k-1)}{T}\left[\tfrac{\text{rad}}{\text{sec}}\right]$.} 
	\label{fig:spectraAR1}
\end{figure}
\begin{table}
	\begin{center}
		\begin{tabular}{|c|c|c|}
			\hline
			\multicolumn{3}{|c|}{\textbf{AR(1) Parameters}} \\
			\hline
			$\us_1$ & $\us_2$ & $\us_3$ \\
			\hline
			$0.84$ & $0.21$ & $-0.57$ \\
			\hline
		\end{tabular}
	\end{center}
	\caption{AR(1) parameters of the sources.}
	\label{table:AR1paramters}
	\vspace{-0.5cm}
\end{table}
First, we consider the case of $\A\in\Rset^{4\times 3}$, where the $M=3$ sources are all Gaussian AR processes of order 1 (AR(1)), each with unit variance and an AR parameter as presented in Table \ref{table:AR1paramters}, with a resulting spectrum as presented in Fig. \ref{fig:spectraAR1}. In the first part of this experiment, the noise level is set to be equal in all $L=4$ sensors, i.e., $\mLambda=\sigma_v^2\I_L$ such that $\sigma_{v_\ell}^2=\sigma_v^2$ for all $1\leq\ell\leq4$. The elements of the mixing matrix
\begin{equation*}\label{exp1matrixA}
\A=\begin{bmatrix}
0.9202 & -0.3396 & 0.8531 \\
0.6021 & -0.7977 & 0.2639 \\
-0.0648 & -0.3944 & -0.0117 \\
0.3877 & -0.5301 & -0.5394
\end{bmatrix},
\end{equation*}
were drawn (once) independently from a standard Normal distribution, $\sigma_v^2$ was set to $0.001$ (an SNR of $30$[dB]) and the sample size was set to $T=1000$.
\begin{figure*}[t]
	\centering
	\includegraphics[width=\textwidth]{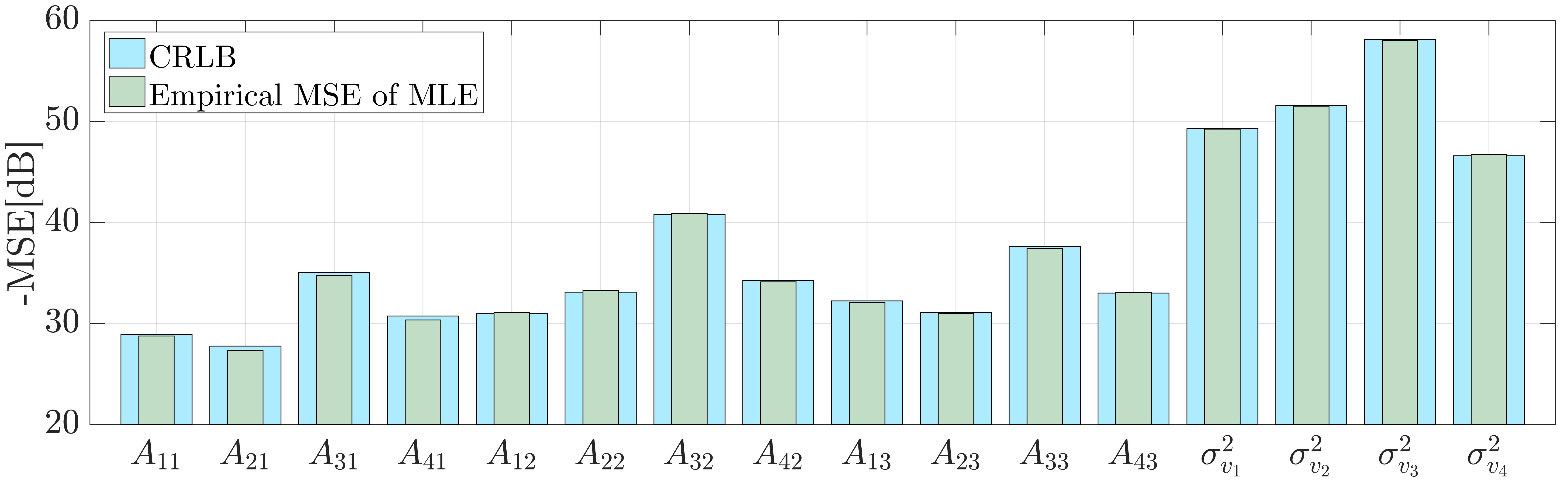}
	\caption{Experiment 1, part 1: CRLB and empirical MSEs of the elements of $\A$ and $\ulambda$, with $T=1000$ and $\sigma_v^2=0.001$. Here, the largest deviation of an empirical result from its corresponding CRLB is $\sim0.4$[dB]. Empirical Results were obtained by $1000$ independent trials.} 
	\label{fig:longsimul_A4X3_T1e3_260218_SNR20dBbetaCRLB}
	\vspace{-0.3cm}
\end{figure*}
We start by empirical cross-verification of the CRLB on the MSE in unbiased joint estimation of $\A$ and $\ulambda$ and their MLEs, focusing on the first phase of our separation-estimation scheme. In addition, although in general the Fisher scoring algorithm is not guaranteed to converge to the MLE, we will also show empirically (in all three experiments) that in our problem it converges to the MLE with high probability. 
The initial solutions for the Fisher scoring algorithm in all three experiments were set as follows. The (fixed) matrix $\A_0=\left[\I_M \; \textrm{O}\right]^{\tps}\in\Rset^{4\times 3}$ was set as the initial solution for the mixing matrix. The noises' variances initial solutions were all set to the smallest eigenvalue of the matrix $\tfrac{1}{T}\X\X^{\tps}$. Now, in order to cope with the sign ambiguity for the performance assessments, we assume that the correct sign of each column of $\A$ is known and compute the empirical estimation error by $\widetilde{\meps}_{A}\triangleq\mathbf{\Xi}\odot\hA_{\text{\tiny{ML}}}-\A$, where $\Xi_{\ell m}\triangleq\text{sign}\left((\hat{A}_{\text{\tiny{ML}}})_{1m}\cdot A_{1m}\right)$ for all $\ell\in\{1,\ldots,L\}$ and $m\in\{1,\ldots,M\}$. In addition, we take $\mathbf{\Xi}\odot\hA_{\text{\tiny{ML}}}$ as the estimated mixing matrix, for the empirical computation of the resulting MSE of the estimated sources. Note that in a real scenario, the estimated sources would be separated exactly to the same extent, only with the true signs remaining unknowns (which does not affect the separation performance), unless some other prior knowledge, which can resolve this ambiguity, is known (as in experiment 3). Fig. \ref{fig:longsimul_A4X3_T1e3_260218_SNR20dBbetaCRLB} presents the theoretical CRLB and the empirical MSEs obtained by the MLEs (computed via the Fisher scoring algorithm) for all the elements of $\A$ and $\ulambda$. As seen from the figure, in these asymptotic conditions the MLEs attain the bound, thus corroborating our derivation for the likelihood equations, the FIM elements and the CRLB.

Next, in Table \ref{table:AverageMSE} we compare the average MSE and the (oracle) MMSE bound. Note that in this scenario, where the SNR is $30[\text{dB}]$, the ML-based MMSE estimates' MSE is, in average, between $\sim1-5[\text{dB}]$ from the MMSE bound. In the next experiment, we shall consider a similar scenario but with lower SNRs, where we expect the optimality gap to become negligible, i.e.,  $\Eset\left[\ud\ud^{\tps}\right]\ll\Eset\left[\uepsilon_{\text{\tiny{MMSE}}}\uepsilon_{\text{\tiny{MMSE}}}^{\tps}\right]$.

In the second part of this experiment, we set
\begin{equation}
	\sigma_{v_\ell}^2=-5(3+\ell)\text{[dB]}, \; 1\leq\ell\leq 4,
\end{equation}
and vary the sample size $T$, so as to examine both a scenario with a different SNR in every sensor and the asymptotic optimality w.r.t. the MMSE bound. As seen from Fig. \ref{fig:For_paper_4x3_diffSNRs_Tvaries_MSE_vs_T}, the asymptotic optimality is evident and the empirical MSEs exhibit a convergence trend towards the MMSE bounds.
\begin{figure}
	\centering
	\includegraphics[width=0.48\textwidth]{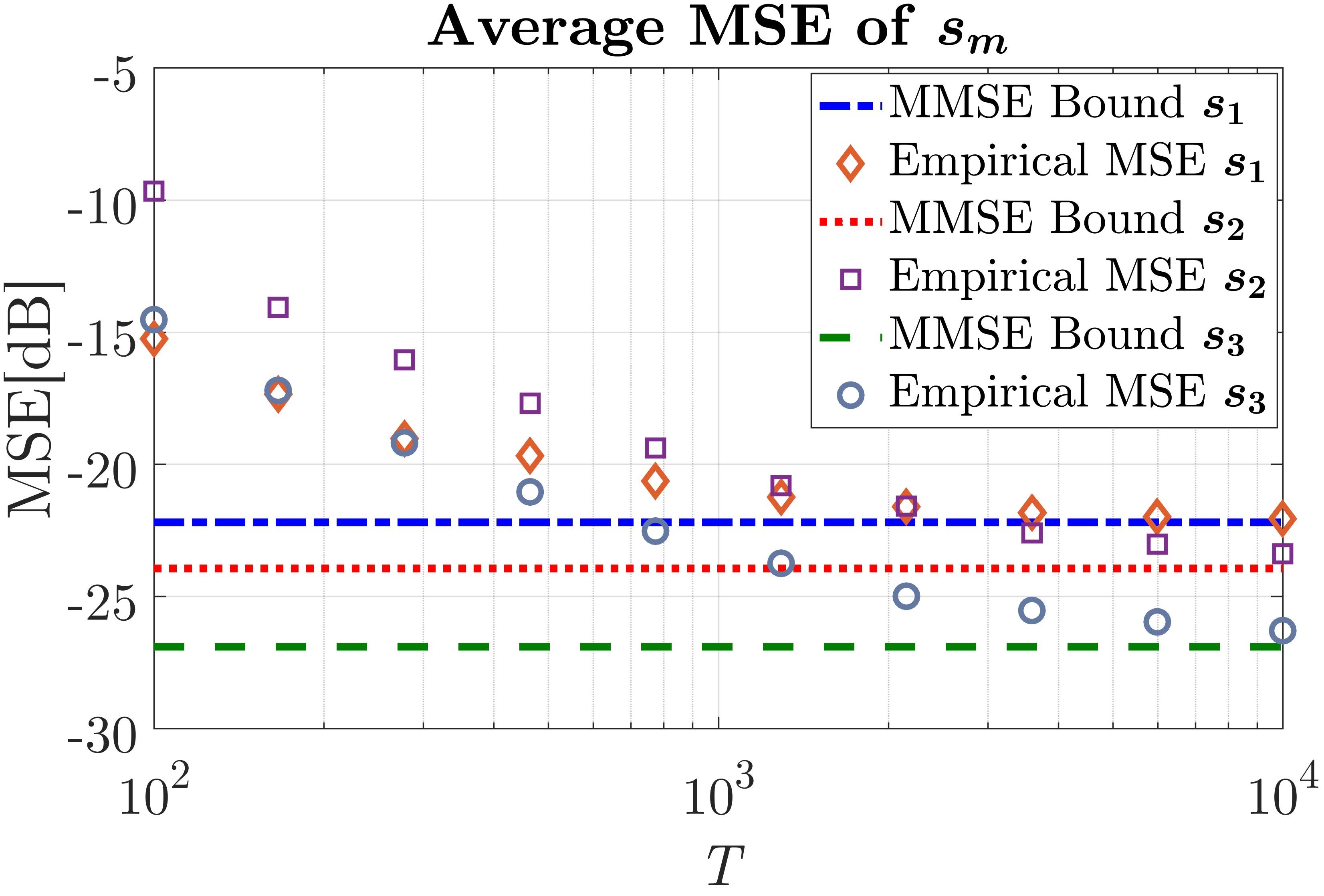}
	\caption{Experiment 1, part 2: MMSE bound and empirical MSEs of the estimated sources vs. sample size $T$, with $\sigma_{v_1}^2=-20$[dB], $\sigma_{v_2}^2=-25$[dB], $\sigma_{v_3}^2=-30$[dB] and $\sigma_{v_4}^2=-35$[dB]. Empirical results were obtained by averaging $100$ independent trials, such that each empirical value in the figure is obtained from an average of (at least) $10^4$ empirical squared errors.} 
	\label{fig:For_paper_4x3_diffSNRs_Tvaries_MSE_vs_T}
	\vspace{-0.2cm}
\end{figure}
\begin{figure*}[t]
	\centering
	\includegraphics[width=\textwidth]{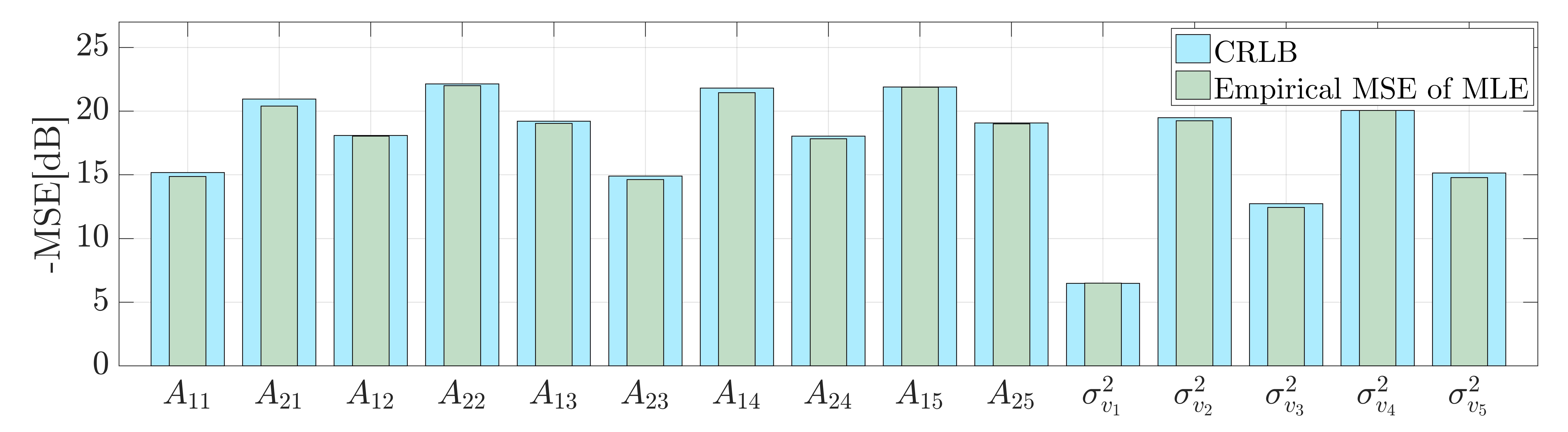}
	\caption{Experiment 2: CRLB and empirical MSEs of the elements of $\A$ and $\ulambda$, with $T=250$ and $\sigma_v^2=1$. Here, the largest deviation of an empirical result from its corresponding CRLB is $\sim0.5$[dB]. Results were obtained by $1000$ independent trials.} 
	\label{fig:For_paper_5x2_0dBSNR_T250_CRLB_MLE}
	\vspace{-0.2cm}
\end{figure*}
\vspace{-0.2cm}
\subsection{Experiment 2: AR Sources in Low SNR}
In this experiment we consider the case of $M=2$ sources (sources $2$ and $3$ from experiment 1) and $L=5$ sensors. Again, the elements of the mixing matrix 
\begin{equation*}\label{exp2matrixA}
\A=\begin{bmatrix}
-0.7270 & -2.1943\\
-0.0249 & 0.8741\\
-1.2327 & 0.8559\\
0.5638 & 0.0343\\
1.0297 & -0.7223
\end{bmatrix},
\end{equation*}
were drawn (once) as in the previous scenario, but the common noise level across all sensors $\sigma_v^2$ (similarly to experiment 1) was set to $1$ , i.e., an SNR level of $0[\text{dB}]$, and the sample size was set to $T=250$.

Fig. \ref{fig:For_paper_5x2_0dBSNR_T250_CRLB_MLE} presents the CRLB and empirical MSEs in ML estimation of $\A$ and $\ulambda$, which demonstrates that the empirical MSEs obtained by the MLEs are close to their corresponding CRLB, again. However, notice that in this experiment, the number of samples per unknown parameter is only a third from the number of samples per unknown parameter considered in the first part of experiment 1 and the SNR is lower by $30$[dB]. Nonetheless, it is seen that the MLEs are quite accurate even in a noisy environment with a relatively small number of samples per unknown parameter. Table \ref{table:AverageMSEexp2} compares the average MSE and the MMSE bound. It is seen that in the low SNR regime the optimality gap (which is in this scenario $\sim0.3$[dB]) becomes negligible w.r.t. the MSE, such that $\Eset\left[\uepsilon_{\text{\tiny{ML-MMSE}}}\uepsilon^{\tps}_{\text{\tiny{ML-MMSE}}}\right]\approx\Eset\left[\uepsilon_{\text{\tiny{MMSE}}}\uepsilon^{\tps}_{\text{\tiny{MMSE}}}\right]$, as expected. Thus, in the low SNR regime the ML-based MMSE sources' estimates are approximately optimal, in the sense of MMSE, even for a relatively small number of samples per unknown parameter.
\begin{table}[t]
	\begin{center}
		\begin{tabular}{|c|c|c|}
			\hline
			\textbf{Source} & \textbf{MMSE Bound [dB]} & \textbf{Average MSE [dB]} \\
			\hline
			$\us_1$ & $-24.34$ & $-22.69$\\
			\hline
			$\us_2$ & $-25.53$ & $-19.67$\\
			\hline
			$\us_3$ & $-26.98$ & $-23.24$\\
			\hline
		\end{tabular}
	\end{center}
	\caption{Experiment 1, part 1: MMSE bound and empirical average MSE.}
	\label{table:AverageMSE}
	\vspace{-0.4cm}
\end{table}
\vspace{-0.25cm}
\subsection{Experiment 3: Blind Spatial Equalization in VLC MIMO}
In our third experiment we demonstrate how the proposed scheme can be applied in the context of indoor VLC MIMO systems (\cite{tran2014performance,mesleh2010indoor,wang2013analysis,wang2015multiuser}) for joint channel and SNR QML estimation and consequent LMMSE estimation of the transmitted signals. In indoor VLC, Light Emitting Diodes (LEDs) and Photo DIodes (PDIs) act as signal transmitters and receptors, respectively, replacing more complex and expensive transmit/receive Radio Frequency (RF) hardware and antennas in RF wireless communication systems. A few of the key advantages of VLC for indoor communications are availability of visible light spectrum at no cost, no licensing/RF radiation issues, inherent security in closed-room applications and the potential to simultaneously provide both energy efficient lighting as well as high-speed, short-range communication using high-luminance LEDs.

In VLC systems, Intensity-Modulation with Direct-Detection (IM/DD) is typically used because of its simplicity. In IM/DD, information is carried on the \textit{intensity} of emitted light. Therefore, the information (electrical) signals modulating the LEDs are real-valued and non-negative. One approach for VLC IM/DD is to use multiple LEDs and PDIs in MIMO array configurations, which has been extensively addressed in recent years (see \cite{fath2013performance,hsu2016high,burton2014experimental,gupta2018performance} and reference therein). In particular, MIMO On-Off Keying (OOK) (\cite{babar2018unary,ndjiongue2014low}) is one of the most popular transmission schemes in VLC, and in optical IM/DD communication systems in general, since in IM/DD transmission the transmitted signals must be non-negative. Thus, OOK is a natural way of meeting this constraint. Here, we shall not elaborate on the physical aspects of this transmission scheme but rather focus on the receiver's estimation algorithm, given the physical model (e.g., \cite{gupta2018performance}).

Consider an indoor VLC system consisting of a transmitter with two LEDs and a non-imaging type receiver with $L=4$ PDIs. In a non-imaging receiver, the signals received directly by multiple PDIs are processed to recover the information bits. Each LED at the transmitter is intensity modulated, i.e., in a given channel use, a LED is either off (which implies a light intensity of zero) or it emits light with some intensity. In each time frame, $M=2$ statistically independent, unit variance OOK signals of length $T$ are transmitted simultaneously from the two LEDs, where each LED transmits one OOK signal. The matrix $\mS\in\{0,2\}^{2\times T}$ represents the (time-domain) OOK signals, where $S_{mt}=s_m[t]$ denotes the $t$-th sample of the $m$-th OOK signal. Here, we assume that each of the source signals is a ``telegraph" process, defined as
\begin{equation}
	\forall t>0:s_m[t]=\begin{cases}
	s_m[t-1],& \text{w.p.}\;1-\alpha_m\\
	\text{mod}\left(s_m[t-1],2\right),& \text{w.p.}\;\alpha_m
	\end{cases},
\end{equation}
\begin{table}[t]
	\begin{center}
		\begin{tabular}{|c|c|c|}
			\hline
			\textbf{Source} & \textbf{MMSE Bound [dB]} & \textbf{Average MSE [dB]} \\
			\hline
			$\us_1$ & $-6.53$ & $-6.21$\\
			\hline
			$\us_2$ & $-9.36$ & $-9.01$\\
			\hline
		\end{tabular}
	\end{center}
	\caption{Experiment 2: MMSE bound and empirical average MSE.}
	\label{table:AverageMSEexp2}
	\vspace{-0.25cm}
\end{table}
where $\text{mod}(\cdot,\cdot)$ is the modulo operator, $0<\alpha_m<1$ is the probability to switch from one state to the other, and $s_m[0]$ is some (``forgotten", irrelevant) initial condition, for $m\in\{1,2\}$.  It is well known that this telegraph process is an ergodic Markovian process (of order 1), which asymptotically has an AR(1)-shaped spectrum with unit variance and an AR(1) parameter $a_m\triangleq2\alpha_m-1$. Thus, in our experiment we consider a case where $\{\alpha_m\}_{m=1}^2$ are (known) user-selected design parameters (related to some pre-coding scheme) of each of the transmitted OOK signals from each LED. In particular, we consider the case of $\alpha_1=0.25$ and $\alpha_2=0.75$, corresponding to the resulting characterizing AR(1) parameters $a_1=-0.5$ and $a_2=0.5$, respectively. Assuming perfect synchronization, the received signals are modeled according to \eqref{modelequation} (in accordance with the physical model given, e.g., in \cite{gupta2018performance}), where $\A$ is the (spatial) channel matrix and in this experiment we also assume the noise level is equal in all 4 PDIs (which is easily obtained as a particular case of our general derivation). We note that the Additive White Gaussian Noise (AWGN) assumption, which models the sum of different noise contributions in this context (e.g.,{\;}thermal noise and ambient light noise), is also commonly used and widely justified \cite{wang2015multiuser,fahamuel2014improved}. Upon reception, the empirical (row) means of $\X$ are subtracted. Then, QML joint estimation of the channel matrix and the common noise level is applied, followed by QML-based LMMSE estimation of the sources. Finally, the transmitted bits are estimated by a threshold decision (above or below zero) for each sample of the separated-estimated sources\footnote{We note that the simple threshold decision rule is not optimal here, in the sense of minimum error probability, since it does not (necessarily) result in the maximum \textit{a posteriori} estimate of the transmitted sequence. However, it still serves as a valid common ground for comparing the different estimates.}. Notice that here, the estimated signals' scales are irrelevant for this threshold decision. Furthermore, notice that in this application the true signs of (all) the channel matrix' elements are known, since in this context all the elements must be non-negative according to the physical model.
\begin{figure*}[]
	\centering
	\begin{subfigure}[b]{0.49\textwidth}
		\includegraphics[width=\textwidth]{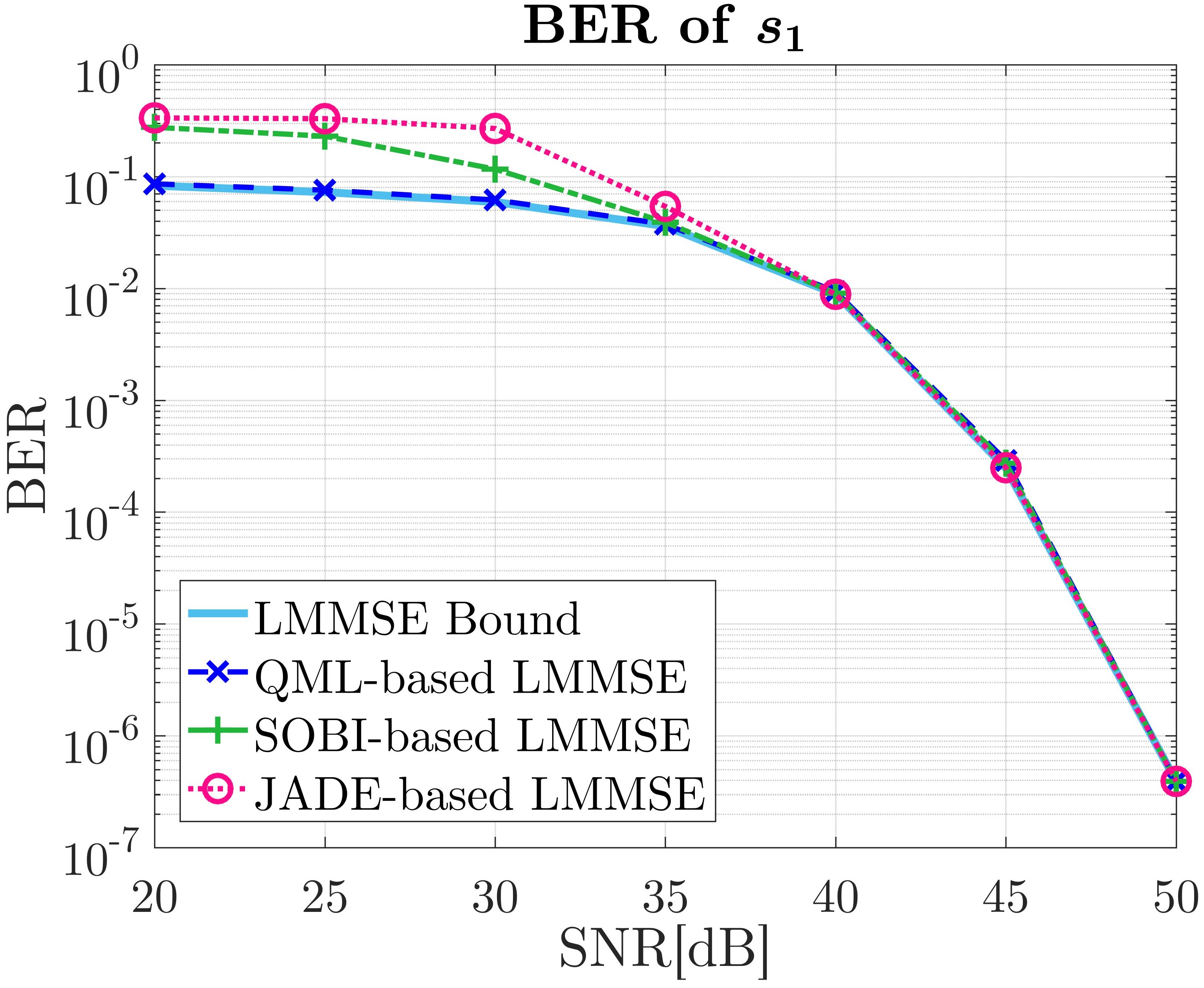}
		\caption{}
		\label{fig:BER_VLC_exp3_s1}
	\end{subfigure}%
	~
	\begin{subfigure}[b]{0.49\textwidth}
		\includegraphics[width=\textwidth]{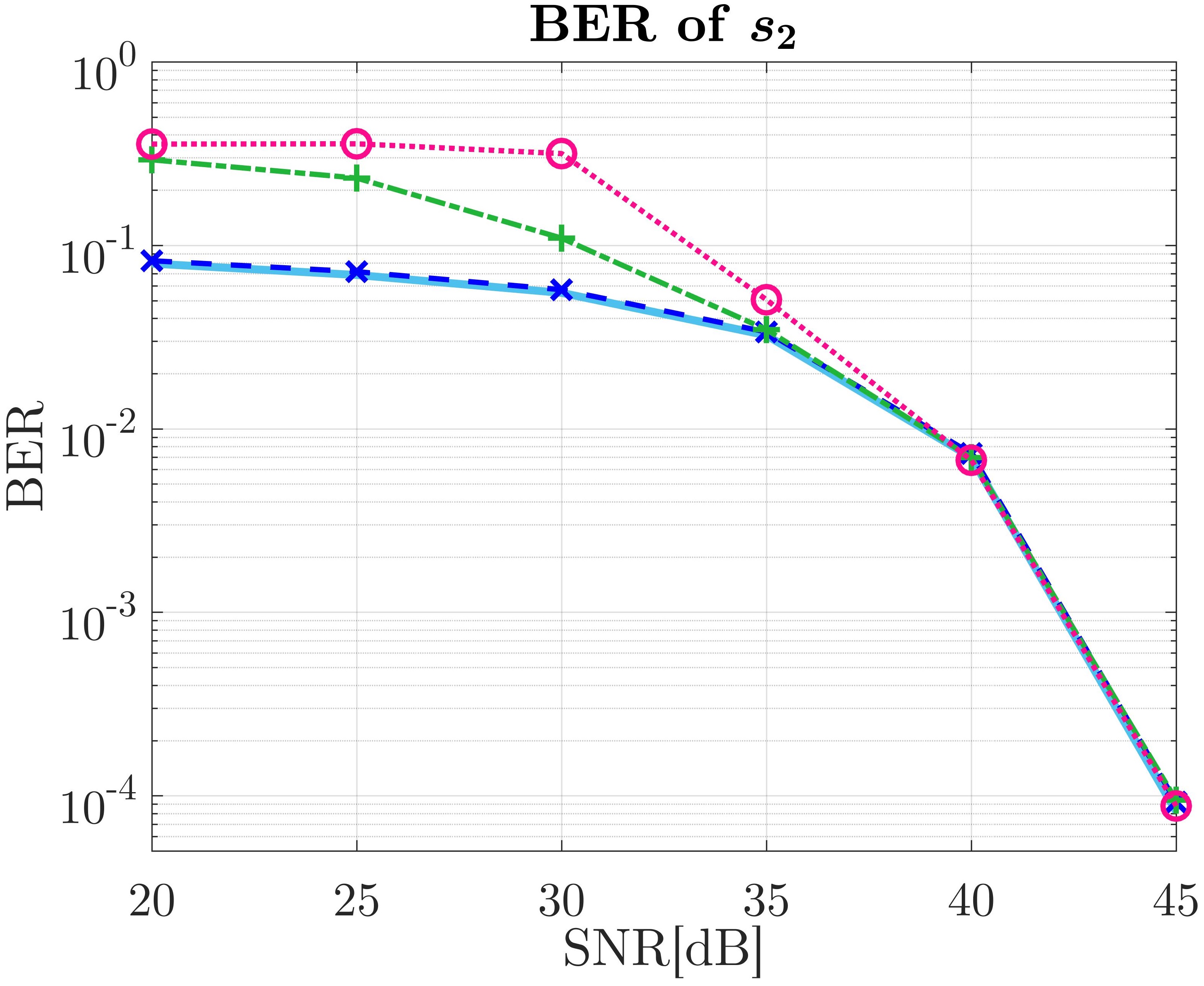}
		\caption{}
		\label{fig:BER_VLC_exp3_s2}
	\end{subfigure}
	\caption{Experiment 3: BER vs. SNR with $T=256$ for OOK sources. Empirical results were obtained by $10^4$ independent trials, corresponding to a total of $\sim10^6$ bits. Above the presented SNR range, zero errors were obtained. (a) BER of source 1 (b) BER of source 2}
	\label{fig:BER_VLC_exp3}
	\vspace{-0.35cm}
\end{figure*}

We compare the performance of the proposed QML-based LMMSE scheme with two similar schemes, replacing the QML phase with the commonly used JADE and SOBI algorithms. In addition, for the JADE- and SOBI-based schemes, we assume that the permutation ambiguity may be solved perfectly by available side-information (not available to the QMLE). The performance is compared in terms of the resulting BER of the two transmitted sequences (sources) for an FFT-compliant size $T=256$. The channel (mixing) matrix
\begin{equation*}\label{exp3matrixA}
\A=\begin{bmatrix}
1.820 & 1.720\\
1.720 & 1.820\\
1.628 & 1.720\\
1.720 & 1.628
\end{bmatrix}\times10^{-6},
\end{equation*}
was taken from \cite{gupta2018performance} (only the relevant first two columns) according to the physical model described therein (see equation (23) and Table I in \cite{gupta2018performance}). Here, we define the SNR as the ratio between the average power of the received (attenuated) sources and the common received noises' power.

As seen from Fig. \ref{fig:BER_VLC_exp3}, when considering all the SNR range, the transmitted bits are best estimated by the QML-based LMMSE, which effectively attains the (oracle) LMMSE bound (an estimate based on the true values of $\A$ and $\sigma_v^2$) throughout all the SNR range. As expected, all three methods perform similarly in the high SNR regime. Notice that although we have shown in subsection \ref{subsec:QMLbasedLMMSE} that the QML-based LMMSE estimate is near-optimal only asymptotically, evidently, in practice it exhibits near-optimal performance even for a reasonable (implementable) value of $T$.
\vspace{-0.2cm}
\section{Conclusion}\label{sec:conclusion}
We presented a comprehensive solution for the separation and estimation of stationary sources from mixtures contaminated by AWGN, based on prior knowledge of the sources' spectra. For Gaussian sources, the solution takes the form of the ML-based MMSE estimate, which asymptotically converges to the oracle MMSE estimate of the sources. As a result, the proposed estimate asymptotically attains the global (oracle) MMSE bound, which bounds the MSE of any estimate for this problem. In the context of the first phase of the proposed scheme  - ML estimation - we provided the CRLB on the MSE of any unbiased estimates of the unknown model parameters and proposed an iterative solution algorithm for computation of the MLEs thereof, which was empirically demonstrated to be an effective solution. We also presented an efficient computation of the sources' ML-based MMSE estimates based on the stationarity of the signals and on the previously obtained MLEs of the model parameters. A qualitative analysis of the estimate's MSE was presented w.r.t. the sample size and the SNR, and all the analytical results were supported by empirical simulation results.

For non-Gaussian sources, the proposed solution takes the form of the Gaussian QML-based LMMSE. This estimate is based on the Gaussian QMLE, which was shown analytically to be a consistent estimate of the model parameters. Consequently, and regardless of the sources' true distributions (beyond their SOS), this estimate is asymptotically sub-optimal, in the sense that it attains the minimal attainable MSE of any linear estimate of the sources. The QML-based LMMSE approach was examined in a simulation experiment of a realistic VLC-MIMO system, for spatial blind equalization and estimation of the transmitted bits sequences, outperforming the JADE- and SOBI-based (pseudo-) LMMSE estimates, while demonstrating how partial \textit{a-priori} information (usually available in communication systems) can be exploited to achieve (sub-)optimal performance.
\vspace{-0.2cm}
\section{Acknowledgment}\label{sec:acknowledgment}
The first author wishes to thank The Yitzhak and Chaya Weinstein Research Institute for Signal Processing for a fellowship.

\bibliography{Bibfile}
\bibliographystyle{unsrt}

\end{document}